\documentstyle[12pt]{article}
\newcommand{\fslash}[1]{\mbox{$\!\not\!#1$}}
\newcommand{\Lag}{{\cal L}}

\newcommand{\be}{\begin{equation}}
\newcommand{\ee}{\end{equation}}

\newcommand{\bold}[1]{\mbox{\boldmath ${#1}$}}

\begin{document}

\baselineskip 4 ex

\title{Axial vector diquark correlations in the nucleon: \\ 
Structure functions and static properties
\footnote{Correspondence to: H. Mineo, E-mail: mineo@nt.phys.s.u-
tokyo.ac.jp}}
       
\author{H. Mineo \\
        Department of Physics, Faculty of Science, University of Tokyo 
\\
        Bunkyo-ku, Hongo, Tokyo 113-0033, Japan \\
{ } \\
       W. Bentz \\
       Department of Physics, School of Science, Tokai University \\
       Hiratsuka-shi, Kanagawa 259-1292, Japan \\
{ } \\
       N. Ishii \\
       The Institute of Physical and Chemical Research (RIKEN) \\
       Hirosawa, Wako-shi, Saitama 351-0198, Japan \\      
{ } \\
       K. Yazaki \\
       Department of Physics, Faculty of Humanities and Sciences, \\
       Tokyo Woman's Christian University \\
       Suginami-ku, Tokyo 167-8585, Japan}

\date{ }
\maketitle
\newpage
\begin{abstract}
In order to extract information on the strength of quark-quark 
correlations in
the axial vector (a.v.) diquark channel      
($J^P=1^+, T=1$), we analyze the quark light cone 
momentum distributions in the nucleon, in particular their flavor 
dependencies, and
the static properties of the nucleon. To construct the nucleon as a 
relativistic 3-quark
bound state, we use a simple 'static' approximation
to the full Faddeev equation in the Nambu-Jona-Lasinio model, including 
correlations in the
scalar ($J^P=0^+, T=0$) and a.v.  diquark
channels. It is shown that the a.v. diquark correlations should be 
rather weak
compared to the scalar ones. From our analysis we extract information 
on the
strength of the correlations as well as on the probability of the a.v. 
diquark channel.  

{\footnotesize PACS numbers: 12.39-x, 12.39.Ki, 14.20.Dh\\
        {\em Keywords}: Structure functions, Effective quark theories, 
Diquark correlations}
\end{abstract}

\newpage

\section{Introduction}
\setcounter{equation}{0}
The investigation of nucleon structure functions and the study of the 
associated quark and gluon light cone (LC) momentum and spin 
distributions
is currently a very active field both experimentally \cite{EXP,SLAC,NM} 
and 
theoretically \cite{BOOK}.
By applying perturbative QCD and factorization theorems \cite{TH} to 
the analysis
of deep inelastic lepton-nucleon scattering experiments, detailed 
information on the quark and gluon distributions in the nucleon
have been (and will further be) extracted. Many excellent 
parameterizations
of quark and gluon distributions now exist \cite{PAR}. Since these 
distributions reflect the
non-perturbative aspects of the problem, at present they cannot be calculated 
directly
from QCD. Effective models based on QCD are powerful tools to extract 
information 
on the effective quark-quark interactions by comparing the calculated
distributions with the empirical ones.

In this paper we will concentrate on the quark LC momentum
distributions relevant for the unpolarized structure functions. 
A particularly interesting feature of the valence ($v$) quark 
distributions
in the proton
is their flavor dependence: The empirical distributions clearly indicate
that $d_v(x)$ is ``softer'' than $u_v(x)$, i.e., it is more 
concentrated in
the region of smaller Bjorken $x$. There is a very simple argument \cite
{FLAV} which
shows that this feature can be naturally explained as a consequence of
the quark-quark (diquark) correlations in the scalar ($J^P=0^+, T=0$) 
channel:
Strong correlations in this channel lead to a ``scalar diquark'' with 
mass
well below twice the constituent quark mass. Since the fraction of the 
LC
momentum carried by the diquark decreases as its mass decreases, and a 
scalar
diquark consists of a valence up and down quark, this means that $d_v$ 
in the
proton is softer than $u_v$ for sufficiently strong scalar diquark 
correlations. 

However, the correlations in the axial vector (a.v.) diquark channel 
($J^P=1^+, T=1$) are
expected to be of some importance, too, from the analogy of non-
relativistic 3-body
calculations.
Compared to a purely quark - scalar diquark model, these correlations 
tend to 
work towards more symmetric valence quark
distributions, since their presence requires a reduction of the 
correlations 
in the scalar diquark channel in order to get the same nucleon mass.
Indeed, the ratio $F_{2n}/F_{2p}$ of the neutron and proton structure
functions in the region $x\rightarrow 1$, where the valence quarks are
dominant, indicates \cite{MEL} that $d_v/u_v$ is probably nonzero (but  
less than the value $\frac{1}{2}$ corresponding to flavor symmetric 
distributions) 
as $x\rightarrow 1$, although there is some model dependence in 
extracting $F_{2n}$ 
from the deuteron data. One of the questions which we want to address
in this paper is therefore how the flavor asymmetry of the valence 
quark distributions
is influenced by the a.v. diquark correlations. 

Another interesting
feature of the distribution functions is the flavor asymmetry of the
antiquark distributions in the proton \cite{KUM}, i.e., 
$\overline{d}(x)>\overline{u}(x)$ over the whole region of Bjorken $x$, 
which leads to the
famous violation of the Gottfried sum rule. A simple and natural 
explanation
of this asymmetry is due to the pion cloud around the valence quarks, 
and
there have been many investigations of this effect based on the familiar
convolution formalism \cite{CONV}.

The primary aim of this paper is to investigate the connection
between the flavor dependence of the valence quark distributions
and the strength of the quark-quark interactions in the a.v. diquark
channel in a model calculation. For this purpose we use the Nambu-Jona-
Lasinio 
\cite{NJL} (NJL) model as an
effective quark theory in the low energy region \cite{NJL1}. Our 
approach is
based on the relativistic Faddeev method \cite{FAD1}, which has been 
used in
many recent works as a powerful tool to investigate the properties
of hadrons \cite{FAD2,FAD3,ISH}. \footnote{Very similar in spirit are 
the covariant quark-diquark 
models \cite{QD,QD1}, which have been extensively used recently to 
describe many properties
of the nucleon like electromagnetic form factors and static 
properties.}  
Although the relativistic
Faddeev equation in the NJL model has been solved exactly including 
also the a.v.
diquark channel \cite{FAD2}, this has not yet been done with LC 
variables,
mainly due to technical reasons associated with regularization and 
angular
momentum projection \cite{LC}. 
Therefore, in this work we consider a simple approximation to the
full Faddeev equation \cite{STAT} (the ``static approximation''), which 
consists in 
neglecting the momentum dependence of the quark exchange kernel.
In this case, the resulting quark-diquark equation can be solved almost
analytically. In a previous work we applied this method to the 
calculation
of structure functions by keeping only the scalar diquark channel \cite
{MIN}, and
here we extend these calculations to include also the a.v. diquark 
channel.
We will derive an upper limit for the strength of the quark-quark 
interaction
in this channel, which can be translated into an upper limit of the
probability (defined here via the contribution to the baryon number).
In order to see whether this result derived from the flavor dependence
of the structure functions is consistent with other properties of the
nucleon or not, we will also investigate the dependence of static 
properties, 
like magnetic moments or the axial vector coupling constant, on the
correlations in the a.v. diquark channel. We will show that a consistent
picture emerges if the probability of the a.v. diquark channel is  
$\stackrel{<}{\sim}10\,\%$, 
although the model has still to be refined for more quantitative 
purposes.

The rest of this paper is organized as follows: In sect. 2 we explain
the model and the form of the nucleon wave function including the
a.v. diquark channel. In sect. 3 we explain our method to calculate
the quark distribution functions, and in sect. 4 we consider the
static properties of the nucleon within the same framework. Sect. 5
is devoted to the discussion of our results, and in sect. 6 we present
a summary and conclusions. 

\section{The model and vertex functions for baryon states}

\setcounter{equation}{0}
We consider SU(2)$_f$ quark Lagrangians of the NJL \cite{NJL} type 
$\Lag = {\bar \psi}
(i\fslash{\partial}-m)\psi +\Lag_I$, where $m$ is the u, d current 
quark mass
and $\Lag_I$ is a chirally symmetric 4-fermi contact interaction.
By applying Fierz transformations, any
$\Lag_I$ can be decomposed into various $q{\overline q}$ and $qq$ 
channels \cite{FAD2}.
The terms relevant for our discussions are given as follows:
\begin{eqnarray}
\Lag_{I,\pi} &=& \frac12 G_{\pi} \left[({\bar \psi}\psi)^2 - ({\bar 
\psi}
\gamma_5 \mbox{\boldmath $\tau$}  \psi)^2 \right] , \label{lpi}\\
{\cal L}_{I,s} &=& G_{s} \left[ {\bar \psi}(\gamma_{5} C ) \tau_2
\beta^A {\bar \psi}^{T}\right]
\left[ \psi^T (C^{-1} \gamma_5 )\tau_2 \beta^{A}\psi \right] , \label
{ls}\\
\Lag_{I,a} &=& G_a \left[ {\bar \psi}(\gamma_{\mu} C )
\mbox{\boldmath $\tau$}\tau_2
\beta^A {\bar \psi}^{T}\right] \cdot
\left[ \psi^T (C^{-1}\gamma_{\mu})
\mbox{\boldmath $\tau$}\tau_2 \beta^{A} \psi\right]  \label{la},
\end{eqnarray}
where the color matrix $\beta^A =\sqrt{\frac{3}{2}} \lambda^A$ 
(A=2,5,7) corresponds to the color
$\overline{3}$ states, and $C=i\gamma_2 \gamma_0$. $\Lag_{I,\pi}$ 
represents the interaction
in the $0^+$ and $0^-$ $q\overline{q}$ channels corresponding to the 
sigma meson and the pion, and 
$\Lag_{I,s}$ ($\Lag_{I,a}$) is the interaction in the $0^+$ ($1^+$) 
$qq$ channel corresponding to 
the scalar diquark (a.v. diquark). The interactions (\ref{lpi}) and 
(\ref{ls}) are invariant
under chiral SU(2)$\times$SU(2) transformations, but (\ref{la}) should 
in principle 
be supplemented by the
vector diquark ($1^-$) interaction term to form a chiral invariant 
lagrangian \cite{ISH}. However, from the naive 
non-relativistic analogy the vector diquark is expected to be of minor 
importance since
it is a $\ell=1$ pair, and will be neglected here.  

The coupling constants $G_{\pi}, G_s$ and $G_a $ are related numerically
to the ones of the original $\Lag_I$. Since $G_{\pi}$ will be 
determined by the properties of the
pion, it is convenient to introduce the ratios
\begin{equation}
r_s = G_s /G_{\pi},\quad r_a = G_a /G_{\pi}.
\end{equation}
These ratios will be treated as parameters reflecting the form of the 
original $\Lag_I$.
(For example, for the 'color current' type interaction lagrangian \cite
{STAT} one has 
$r_s=0.5$ and $r_a=0.25$.) 

The reduced t-matrices in the pionic $q{\overline q}$ channel, and the scalar and 
axial vector $qq$ channels are obtained 
by solving the Bethe-Salpeter equations in these channels, and the 
results are the following
standard NJL model expressions \cite{NJL,NJL1}:
\begin{eqnarray}
\tau_{\pi}(k) = \frac{-2iG_{\pi}}{1+2G_{\pi}\Pi_{\pi} (k^2 )},\quad
\tau_s (k) = \frac{4iG_s}{1+2G_s\Pi_s  (k^2 )}, \label{tau_s} \\
\tau_a^{\mu\nu} (k) = 4iG_a \left[ g^{\mu\nu}-
\frac{2G_a \Pi_a (k^2 )}{1+2G_a \Pi_a (k^2 )}
\left(g^{\mu\nu}-\frac{k^{\mu}k^{\nu}}{k^2 }\right) \right],
\label{tau_a}
\end{eqnarray}
with the ``bubble graphs''
\be
\Pi_{\pi}(k^2 ) = \Pi_s (k^2 )=6i\int\frac{d^4 q}{(2\pi)^4 }\mbox{tr}_D
[\gamma_5 S(q)\gamma_5 S(k+q)] \label{pis},
\ee
\be
{\Pi_a} (k^2 )\left(g^{\mu\nu}-\frac{k^{\mu}k^{\nu}}{k^2 }\right)=
6i\int \frac{d^4 q}{(2\pi)^4 }\mbox{tr}_D
[\gamma^{\mu} S(q)\gamma^{\nu} S(k+q)],
\label{pia}
\ee
where $S(k)=(\fslash{k}-M+i\epsilon)^{-1}$ is the quark Feynman 
propagator
and $M$ is the constituent quark mass, which is related to the current 
quark mass $m$ via
the familiar gap equation of the NJL model \cite{NJL}.

For the evaluation of these and other loop integrals we will
use the ``covariant three-momentum cut-off scheme'' \cite{NJL1,LC}, by 
which we mean the
following: After parameterizing the expressions for the Green functions 
in
terms of Lorentz invariant functions (like $\Pi_a$ in eq. (\ref{pia})), 
these Lorentz
invariants are calculated by introducing a sharp three-momentum cut-off 
($\Lambda_3$) in a
particular Lorentz frame, which we take to be the rest frame (${\bold k}
=0$) 
in case of two-point functions and the Breit frame in the case of three- 
point functions involving one vertex for the external field. 
The Lorentz invariant 
expressions are then recovered by ``boosting'' to a general frame. (In 
the case of the above 
bubble graphs, this ``boosting'' simply means the replacement $k_0^2
\rightarrow k^2$.) 
The value of $\Lambda_3$ is assumed to be the same for all loop 
integrals. The three-momentum cut-off scheme used here preserves
the conservation of electric charge and baryon number, but {\em not}
the full current conservation which involves finite momentum transfer
by the external field. It also preserves low energy theorems based on 
chiral symmetry, but not the full PCAC and Goldberger-Treiman relation
(see sect. 5.2). 

In order to obtain the nucleon vertex function, we solve the homogeneous
Faddeev equation in the  ``static approximation'' \cite{STAT,MIN}, 
where the momentum
dependence of the quark exchange kernel ($Z$) is neglected. Performing 
the projections to
color singlet and isospin $\frac12$ as in \cite{FAD2}, the resulting
equation for the baryon vertex function $\Gamma(p)$ becomes
\begin{equation}
\Gamma^{a} (p) = Z^{aa'} \Pi_N^{a'b}(p)\Gamma^{b}(p) \equiv K^{ab}(p)
\Gamma^{b}(p),
\label{Faddeev}
\end{equation}
where $Z$ is the quark exchange kernel in the static approximation,
\begin{equation}
Z^{aa'} = \frac{3}{M} \left(
\begin{array}{cc}
1 & \sqrt{3}\gamma^{\mu '}\gamma_5 \\
\sqrt{3}\gamma_5 \gamma^{\mu} & -\gamma^{\mu'}\gamma^{\mu} \\
\end{array}\right), \label{exk} 
\end{equation}
and $\Pi_N(p)$ is the quark-diquark bubble graph
\begin{eqnarray}
\Pi_N^{a'b}(p) &=& \int \frac{d^4 k}{(2\pi)^4 } \tau^{a'b}(k) S(p-k) 
\label{pin} \\
\tau^{a'b}(k) &=& \left(
\begin{array}{cc}
\tau_s(k) & 0 \\
0 & \tau_a^{\mu'\nu}(k) \\
\end{array}\right) \equiv  
\mbox{diag}\,(\tau_s (k), \tau_a^{\mu'\nu}(k)). \label
{taun}
\end{eqnarray}
In the above equations, roman indices ($a,a',\dots$) combine one index 
for 
the scalar diquark (denoted as 5) with the 4 Lorentz indices 
($\mu,\mu',\dots)$ for the 
a.v. diquark.
Together with the Dirac index for the quark (not shown explicitly), the 
kernel $K$ in
eq.(\ref{Faddeev}) is a $20\times 20$ matrix, which can be reduced 
to $10\times 10$ by
projection to positive parity states (see Appendix A). 
We directly diagonalize this kernel
in the rest frame of the baryon (${\bold p}=0$) by unitary 
transformations in Appendix A. 
It separates into a $6\times 6$ block corresponding to spin $\frac{1}{2}
$ states (nucleon)
\footnote{The corresponding 6 basis states correspond to the coupling 
of the quark with a scalar diquark, with the 
time component of the a.v.  diquark, and with the space components of 
the a.v. diquark, where
each of these states has spin degeneracy 2.}, 
and a $4\times 4$ block for the spin $\frac{3}{2}$ states. 
After boosting the eigenfunction to a general frame, we obtain for the 
nucleon state
$\Gamma^a_N=\left(\Gamma^5_N, \Gamma^{\mu}_N\right)$, where 
\begin{eqnarray}
\Gamma_N^5(p,s) &=& \alpha_1 u_N(p,s) \label{wf1s} \\
\Gamma_N^{\mu}(p,s) &=& \alpha_2 \frac{p^{\mu}}{M_N} \gamma_5 u_N(p,s) 
+ \alpha_3 \sum_{\lambda s'}
\left(1 \frac{1}{2},\lambda s'|\frac{1}{2} s\right) \epsilon_{\lambda}^
{\mu}(p) u_N(p,s'). 
\nonumber \\
\label{wf1v} 
\end{eqnarray}
Here $u_N(p,s)$ is a Dirac spinor, with spin projection $s$, depending on 
the nucleon mass $M_N$ and normalized according to
$\overline{u}_N u_N =1$, and
$\epsilon_{\lambda}^{\mu}(p)$ is the polarization 4-vector, which also 
depends on 
$M_N$, and is obtained as usual by applying a Lorentz boost to $(0,
{\bold \epsilon}_{\lambda})$, where 
${\bold \epsilon}_{\lambda}$ are the spherical
unit vectors with $\lambda=\pm 1, 0$. The homogeneous equation for the 
3 coefficients $\alpha_1, \alpha_2, \alpha_3$
leads to the eigenvalue equation for the nucleon mass $M_N$ (see 
Appendix A). 
Note that in the last term of (\ref{wf1v}) the spin projection $s'$ in 
the Dirac spinor 
refers to the quark. 
The vertex function for
spin $\frac{3}{2}$ is given by an expression similar to the last term 
in  (\ref{wf1v}), but with 
the diquark and quark spins coupled to $\frac{3}{2}$ instead of $\frac
{1}{2}$. This vertex function
is therefore of the standard Rarita-Schwinger form.

By simple manipulations explained in Appendix A, eq.(\ref{wf1v}) can be 
shown to
be equivalent to the following covariant form:
\begin{eqnarray}
\Gamma_N^{\mu}(p,s)= a_2 \frac{p^{\mu}}{M_N} \gamma_5 u_N(p,s) + a_3 
\gamma^{\mu} \gamma_5 u_N(p,s)
\label{wf2}
\end{eqnarray}
with $a_2=\alpha_2 - \alpha_3/\sqrt{3}$ and $a_3=-\alpha_3/\sqrt{3}$. 
The form (\ref{wf2}), which is a special
case of the general covariant decomposition of the baryon vertex 
functions 
given in ref. \cite{QD}, is more convenient
for actual Feynman diagram calculations 
\footnote{The isospin and color parts of the vertex functions 
are standard and not shown explicitly here. We also note that it is of 
course possible to start directly
from the covariant parameterization (\ref{wf2}), together with the 
Rarita-Schwinger type part corresponding to 
spin $\frac{3}{2}$, 
introduce a covariant decomposition of the kernel in (\ref{Faddeev}) 
and derive the homegenous equations for
the coefficients $a_1\equiv \alpha_1, a_2$ and $a_3$.}.    

The normalization condition for the nucleon vertex function, which we 
give here in terms of LC
variables $p_{\pm} = (1/\sqrt{2})(p_0 \pm p_3)$ for later use, is as 
follows \cite{ASA,LC}:
\begin{eqnarray}
\frac{1}{2 p_-} {\bar \Gamma_N^{a} (p)} \frac{\partial\Pi_{N,ab} (p)}
{\partial p_+}
\Gamma_N^{b} (p)= 1 \equiv W_s + W_a. \label{norm}
\end{eqnarray}
Here we introduced the ``weights'' of the scalar and a.v.  
diquark channels, where
$W_s$ is the contribution from $a=b=5$ and $W_a$ is the remaining part. 
In any treatment
which preserved the Ward identities, the l.h.s. of (\ref{norm}) 
corresponds to the matrix element of
the baryon number operator \cite{MIN}, and therefore $W_s$ and $W_a$ 
are the contributions of the scalar and a.v. diquark channels to the 
total baryon number.  

Similarly, the vertex function for the $\Delta$ isobar state is 
obtained from an eigenvalue
equation similar to (\ref{Faddeev}) with the kernel projected to color 
$\overline{3}$ and $T=\frac{3}{2}$:
\begin{eqnarray}
K^{\mu \nu}(p) = \frac{6}{M} \left(\gamma^{\mu} \gamma_{\rho}\right) 
\Pi_N^{\rho \nu}(p)  
\label{kd}
\end{eqnarray}
with the same quark-diquark bubble $\Pi_N(p)$ graph as before (see eq.
(\ref{pin})). 
The corresponding eigenfunction, which
has the same form as the last term in eq. (\ref{wf1v}) but with the 
nucleon mass replaced by the delta mass 
$M_{\Delta}$ and the total spin $\frac{1}{2}$ replaced by $\frac{3}{2}
$  
in the Clebsch-Gordan coefficient, is of the standard Rarita-Schwinger 
form. The eigenvalue equation for
the $M_{\Delta}$ is also given in Appendix A.    

\section{Quark distribution functions}

\setcounter{equation}{0}
The quark LC momentum distribution in the proton is defined as \cite
{JAF}
\be
\tilde{f}_{q/P}(x) = \frac12 \int\frac{dz^- }{2\pi}e^{ip_{-}xz^- } 
\langle 
p\vert T\left({\overline \psi_q}(0)\gamma^+ \psi_q(z^- )\right)\vert p
\rangle, 
\label{dis}
\ee
where $x$ is the fraction of the proton's LC momentum component $p_- $ 
carried by the quark 
with flavor $q=u,d$, 
and $\vert p>$ denotes the proton state with momentum $p$. As explained 
in \cite{LC,MIN}, the evaluation of the
distribution (\ref{dis}) can be reduced to a straightforward Feynman 
diagram calculation by noting that
it is nothing but the Fourier transform of the quark two-point function 
in the 
proton traced with $\gamma^+$, where the
component $k_-$ of the quark LC momentum is fixed as $k_-=x\,p_-$ and 
the other quark momentum
components ($k_+, {\bf k}_{\perp}$) are integrated out. The relevant 
Feynman 
diagrams in our quark-diquark model are shown
in Fig. 1, where the operator insertion stands for $\gamma^+ \delta(k_--
p_-\,x) (1\pm \tau_z)/2$ for the $u$ ($d$) quark distribution.

The isospin matrix elements are easily evaluated, and the distributions 
can be expressed as follows
\footnote{The distribution (\ref{dis}) is nonzero in the interval $-
1<x<1$, and the physical
quark and antiquark distributions are obtained as $f_{q/P}(x)=\tilde{f}_
{q/P}(x)$ and
$f_{\overline{q}/P}(x)=-\tilde{f}_{q/P}(-x)$ for $0<x<1$. We denote the 
quark 
flavor as $Q=U,D$ for the diagrams of Fig.1 to distinguish this 'valence
quark picture' from the case including the pion cloud. In writing down 
the expressions (\ref{genu}) and
(\ref{gend}) we used the fact that there is no mixing between the 
scalar and a.v. diquark
diagrams for the quark distributions.}:   
\begin{eqnarray}
f_{U/P}(x)&=& F_{Q/P}^s (x) + \frac{1}{2} F_{Q(D)/P}^s (x)
+\frac{1}{3}F_{Q/P}^a (x) +\frac{5}{6}F_{Q(D)/P}^a (x)\nonumber \\
\label{genu} \\
f_{D/P}(x)&=& \frac{1}{2} F_{Q(D)/P}^s (x)
+\frac{2}{3}F_{Q/P}^a (x) +\frac{1}{6}F_{Q(D)/P}^a (x). \label{gend} 
\end{eqnarray}
Here the distribution $F_{Q/P}^d \,\,\,(d=s,a)$ corresponds to the 
first diagram of fig.1,
which will be called the ``quark diagram'' and involves the diquark t-
matrix $\tau_d$, 
and $F_{Q(D)/P}^d$ to the second diagram (``diquark diagram'') with two
$\tau_d$ 's. To derive their explicit expressions from the Feynman 
diagrams, one uses
the Ward-like identity 
${\displaystyle S(k)\gamma^+ S(k) =-\frac{\partial}{\partial k_+ } S(k)}
$ for the
constituent quark propagator and performs partial integrations w.r.t. 
$k_+ $, which is
justified since the regularization scheme to be discussed below (Lepage-
Brodsky scheme \cite{LB})
does not restrict the $k_+$ integrals which are convergent. In this way 
we obtain from the 'quark diagram':
\be
F_{Q/P}(x) = \frac{1}{2p_- } {\bar \Gamma_N }^{a} (p)\left( \frac
{\partial }
{\partial p_+}
\Pi_N^{ab} (x,p) \right) \Gamma_N^{b} (p) \equiv F_{Q/P}^s + 
F_{Q/P}^a,  
\label{fq}
\ee
where the quark-diquark bubble for fixed $k_-$ is given by
\be
\Pi_N^{ab} (x,p) = \int \frac{d^4 k}{(2\pi)^4 }\delta ( x-\frac{k_- }
{p_- }
)S(k) \tau^{ab}(p-k) = {\rm diag}\left(\Pi_N^{55}(x,p),\Pi_N^
{\mu\nu}(x,p)\right) 
\label{piq}
\ee
For the evaluation of the 'diquark diagram' of Fig.1 one also uses the 
Ward-like identity
for the quark propagator on which the insertion is made, and introduces 
the fraction of the
nucleon's momentum component $p_-$ carried by the diquark ($y$) and the 
fraction of the 
diquark's momentum component $q_-$ carried by the quark inside the 
diquark ($z$) via the identity
\begin{eqnarray}
1 = \int {\rm d}y\, \int {\rm d}z \delta(y-\frac{q_-}{p_-}) \,\, \delta
(z-\frac{k_-}{q_-}). \nonumber 
\end{eqnarray}
In this way we obtain 
\begin{eqnarray}
F_{Q(D)/P}(x)&=& \frac{-i}{2p_-} \int_0^1 {\rm d} y \int_0^1 {\rm d} z 
\int \frac{{\rm d}^4q}{(2\pi)^4}
\delta(y-\frac{q_-}{p_-}) \nonumber \\
&\times& {\bar \Gamma_N }^a(p) S_F(p-q) \left(\tau_{ab}(q) \frac
{\partial \Pi^{bc}(z,q)}
{\partial q_+} \tau_{cd}(q)\right) {\bar \Gamma_N }^d(p), \label{di1}
\end{eqnarray}
where the quark-quark bubble graph for fixed LC momentum fraction $z$ 
of one of the quarks is
given by ${\displaystyle \Pi^{ab}(z,q) = {\rm diag}\left(\Pi_s(z,q^2), 
\Pi_a^{\mu \nu}(z,q)\right)}$ with
\begin{eqnarray}
\Pi_s (z, q^2 ) &=&
6i\int\frac{d^4 k}{(2\pi)^4 }\delta(z-\frac{k_-}{q_-})\mbox{tr}_D 
 \left(\gamma_5 S(k)\gamma_5 S(q-k)\right) \label{opens} \\
\Pi_a^{\mu\nu}(z,q)&\equiv&\Pi_a(z, q^2) 
\left(g^{\mu\nu}-\frac{q^{\mu}q^{\nu}}{q^2}\right)\nonumber\\
&=& 6i\int \frac{d^4 k}{(2\pi)^4 }
\delta(z- \frac{k_-}{q_-})\mbox{tr}_D 
[\gamma^{\mu} S(k)\gamma^{\nu} S(q-k)]. \label{opena} 
\end{eqnarray}
Since the quark-quark bubble graphs $\Pi^{ab}(z,q)$ and $\Pi^{ab}(q)$ 
have the same structure w.r.t. the 
diquark indices (see (\ref{pia}) and (\ref{opena})), we can define the 
quark 
LC momentum distribution $F_{Q/D}^d (z,q^2)$ within a 
diquark $d=s,a$ with virtuality $q^2$ by
\begin{eqnarray}
\frac{\partial \Pi^{ab}(z,q)} {\partial q_+} \equiv {\rm diag} \left( 
\frac{1}{2} 
\frac{\partial \Pi_s(q^2)}
{\partial q_+} F_{Q/D}^s (z,q^2), \frac{1}{2} \frac{\partial \Pi^{\mu 
\nu}_a(q)}
{\partial q_+} F_{Q/D}^a (z,q^2) \right) \label{defq}
\end{eqnarray}
By taking the trace over the Lorentz indices in this equation for the 
a.v. contribution, we find
explicitly
\begin{eqnarray}
F_{Q/D}^d (z,q^2) &=& -2 g_d^2 (q^2) \frac{\partial \Pi_d(z,q^2)} 
{\partial q^2}, \label{defq1}
\end{eqnarray}
where
\begin{eqnarray}
g_d (q^2) &\equiv& \frac{-1}{\frac{\partial \Pi_d(q^2)}{\partial q^2}} 
\label{coupl}
\end{eqnarray}
with $d=s,a$. The quantities (\ref{coupl}) are the natural 
generalizations of the quark-diquark 
coupling constants to arbitrary virtuality $q^2$. Inserting the 
definition
(\ref{defq}) into (\ref{di1}), using the Ward-like identity for the 
derivatives of quark-quark
bubble graphs
\begin{eqnarray}
\left(\tau_{ab}(q) \frac{\partial \Pi^{bc}(q)}{\partial q_+} \tau_{cd}
(q)\right) = -2i
\frac{\partial \tau^{ad}(q)}{\partial q_+}, \label{ward}
\end{eqnarray}
and performing a partial integration w.r.t. $q_+$ we finally arrive at
\begin{eqnarray}
F_{Q(D)/P}(x) &=& \int_0^1 {\rm d}y \int_0^1 {\rm d}z \delta(x-yz) 
\int _{-\infty}^{\infty} {\rm d}q_0^2
\sum_{d=s,a} F_{Q/D}^d (z,q_0^2)\,F_{D/P}^d (y,q_0^2). \nonumber \\
\label{fd}
\end{eqnarray}  
Here the LC momentum distribution of the diquark d with virtuality 
$q_0^2$ in the proton is given by
\begin{eqnarray}
F_{D/P}(y,q_0^2) &\equiv& F_{D/P}^s(y,q_0^2) + F_{D/P}^a(y,q_0^2) 
\nonumber \\
&=& {\bar \Gamma_N }^{a} (p) \left(\frac{1}{2 p_-} \frac{\partial}
{\partial p_+} + 
y \frac{\partial}{\partial q_0^2}\right) \Pi_{N}^{ab}(y,q_0^2;p) 
\Gamma_N^{b}(p),\nonumber \\
\label{dp} 
\end{eqnarray}   
where the quark-diquark bubble graph for fixed virtuality and LC 
momentum fraction of the diquark
is given by
\begin{eqnarray}
\Pi_N^{ab}(y,q_0^2 ;p)=\int \frac{d^4 q}{(2\pi)^4 }\delta 
(y-\frac{q_- }{p_- })\delta(q^2 -q_0^2 ) S_F(p-q) \tau^{ab}(q). \label
{disd}
\end{eqnarray}
The normalizations of the distributions $F_{Q/P}(x)$ and $F_{Q(D)/P}(x)
$ follow from eqs. (\ref{fq}), (\ref{fd})
and (\ref{norm}) as follows \footnote{Note that the relations $\int_0^1 
{\rm d}z F_{Q/D}^d(z,q_0^2) = 2$ and
$\int_0^1 {\rm d}z\,z F_{Q/D}^d(z,q_0^2) = 1$ hold for any virtuality 
$q_0^2$, and that the second term in (\ref{dp})
gives a vanishing surface term when integrated over $q_0^2$.}:
\begin{eqnarray}
\int_0^1 {\rm d}x F_{Q/P}(x) &=& W_s + W_a = 1 \label{nq} \\
\int_0^1 {\rm d}x F_{Q(D)/P}(x) &=& 2(W_s + W_a) = 2, \label{nd} 
\end{eqnarray}
which lead to the correct number sum rules for the distributions (\ref
{genu}) and (\ref{gend}). In the
same way the validity of the momentum sum rule can also be checked 
analytically. These sum rules hold in
any regularization scheme which does not restrict the LC plus-
components ($k_+$) of the 
loop momenta \cite{MIN}. 

For the evaluation of the quark-quark and quark-diquark bubble graphs 
with fixed LC momentum
fraction of one of the constituents we use the Lepage-Brodsky (LB) 
regularization scheme \cite{LB}, which is
equivalent to the covariant three-momentum cut-off scheme discussed in the 
previous subsection if
all internal momenta are integrated out \cite{LC}. 
Because this scheme does not restrict the LC plus-components of the loop
momenta, it preserves the number and momentum sum rules. 
In practice, to regularize bubble graphs involving Lorentz indices 
like $\Pi_a^{\mu \nu}$ of eq. (\ref{opena}) or $\Pi_N^{\mu \nu}$ of eq. 
(\ref{disd}), one first decomposes
them into a sum of Lorentz tensors multiplied by Lorentz scalar 
quantities (which, for the case of $\Pi_N^{\mu \nu}$,
are Dirac matrices $\propto 1$ or $\fslash{p}$), and then applies the 
LB scheme to
evaluate the scalar functions in the frame where the transverse 
components of the total momentum are zero.
(To extract the scalar functions from the general Lorentz decomposition 
in this frame, one considers appropriate
combinations of the Lorentz indices.) The results are then generalized 
to an arbitrary frame in a similar
way as explained for the covariant 3-momentum cut-off.   

The model described above gives only valence-like distributions at the 
low energy scale. In order to describe
also the sea quark distributions, we take into account the effects of 
the pion cloud around the constituent
quarks in the same way as described in detail in ref. \cite{MIN}. This 
treatment corresponds to the standard
one-dimensional convolution formalism \cite{CONV}, which involves an on-
shell approximation for the ``parent quark''.
The resulting distribution functions including the pion cloud 
contributions are given by
\begin{eqnarray}
f_{q/P}(x) = \sum_{Q=U,D} \int_0^1 {\rm d}y \int_0^1 {\rm d}z \delta(x-
yz) f_{q/Q}(z) f_{Q/P}(y) 
\label{conv}
\end{eqnarray}
and a similar expression with $q\rightarrow \overline{q}$. The valence-
like distributions $f_{Q/P}(y)$ are
given by eq. (\ref{genu}) and (\ref{gend}), and the expressions for the 
quark and antiquark distributions 
within an on-shell parent quark ($f_{q/Q}$ and $f_{\overline{q}/Q}$) 
are given in ref.\cite{MIN} and  represented
by the Feynman diagrams shown in fig.2.

\section{Static properties of the nucleon}

\setcounter{equation}{0}
In this section we give the basic expressions in our quark-diquark 
model for the
magnetic moments, the isovector and isoscalar axial vector coupling 
constants and the pion-
nucleon coupling constant. The Feynman diagrams for the valence quark 
contributions
and the pionic cloud effects are the same as those for the LC momentum 
distributions (see Figs.
1 and 2) with the appropriate operator insertions, where now we have to 
integrate over all four 
components of the loop momenta. The actual evaluation of these Feynman 
diagrams is similar to
the one described in ref. \cite{QD1}.

\subsection{Magnetic moments}
For the electromagnetic current in the valence quark picture, we insert 
the operator $Q_q \gamma^{\mu}$ 
into the diagrams of Fig.1, where $Q_q$ is the charge operator for the 
quark. Separating the isospin
matrix elements, we write for the magnetic moments of the proton and 
neutron:
\begin{eqnarray}
\mu_p &=& \frac23{\cal F}^s_Q + \frac13 {\cal F}^s_D + {\cal F}^a_D
+ \frac{1}{\sqrt3}{\cal F}^m_D  \label{mup}\\
\mu_n &=& -\frac13{\cal F}^s_Q + \frac13 {\cal F}^s_D +
\frac13{\cal F}_Q^a -\frac13 {\cal F}^a_D - \frac{1}{\sqrt3}{\cal F}
^m_D, \label{mun}
\end{eqnarray}
where the subscripts $Q$ and $D$ refer to the quark and diquark 
diagrams of fig. 1, respectively,
and the superscripts $s$, $a$ and $m$ denote the contributions due to 
the scalar diquark,
the a.v. diquark, and the scalar-a.v. mixing terms, respectively. These 
quantities are obtained
by expressing the contribution to the current from the quark diagram 
($j^{\mu}_Q$) 
and the diquark diagram ($j^{\mu}_D$) in terms of Dirac-Pauli form 
factors,
\begin{eqnarray}
j^{\mu}_R(p',p) &=& {\bar \Gamma}_N^a (p') \lambda_{R, ab}^{\mu} (p',p) 
{\bar \Gamma}_N^b(p) \label{jq} \\
&\equiv& \sum_{d=s,a,m} {\bar u}_N(p') \left[{\cal F}_{1R}^d(q^2) 
\gamma^{\mu} + {\cal F}_{2R}^d(q^2) 
\frac{i \sigma^{\mu \nu} q_{\nu}}{2M_N} \right] u_N(p)  \label{par} 
\nonumber
\end{eqnarray}
where $R=Q, D$. There are no mixing terms from the quark current 
diagram 
(${\cal F}_{1Q}^m={\cal F}_{2Q}^m=0$.) 
Then the various contributions to the magnetic moments in eq. (\ref
{mup}) and (\ref{mun})
are obtained by
${\cal F}_Q^d = {\cal F}_{1Q}^d(0) + {\cal F}_{2Q}^d(0)$ ($d=s,a$) and
${\cal F}_D^d = {\cal F}_{1D}^d(0) + {\cal F}_{2D}^d(0)$ ($d=s,a,m$). 

The vertices $\lambda_{Q, ab}^{\mu}$ and $\lambda_{D, ab}^{\mu}$ in 
(\ref{jq})
are obtained from Fig.1 as follows:
\begin{eqnarray}
\lambda^{\mu}_{Q,ab}(p',p) &=& -\int \frac{d^4 k}{(2\pi)^4} S_F(k+q)
\gamma^{\mu}S_F(k)\tau_{ab}(p-k)
\label{emq} \\
\lambda^{\mu}_{D,ad}(p',p)&=&-i \int \frac{d^4 k}{(2\pi)^4}S_F(p-k)\tau_
{ab}(k')
\Lambda^{\mu}_{bc}(k',k)\tau_{cd}(k),
\label{emd}
\end{eqnarray}
where the 3-point function for the electromagnetic current of the 
diquark is given by
\begin{equation}
\Lambda^{\mu}_{ab}(k',k)= 3i \int\frac{d^4 t}{(2\pi)^4 }
\mbox{tr}_D [S(t+q)\gamma^{\mu} S(t) \gamma_b S(t-k)\gamma_a ].
\label{dc}
\end{equation}

The evaluation of the Dirac-Pauli form factors in eq. (\ref{par})
in the covariant three-momentum cut-off scheme proceeds as follows: For 
each term in the loop
integrals (\ref{emq}) and (\ref{emd}) which involves Lorentz tensors 
like
$k^\nu$ or $k^{\mu}k^{\nu}$, a parameterization in terms of Lorentz 
tensors w.r.t. 
the external (fixed)
momenta, multiplied by Lorentz invariant functions, is introduced. In 
this way one derives formal expressions of the
Dirac-Pauli form factors in terms of these Lorentz invariant functions. 
For the evaluation of the
magnetic moments, it is sufficient to expand these functions up to 
${\cal O}(q)$, and the corresponding
coefficients are evaluated by introducing the 3-momentum cut-off in a 
particular frame, 
i.e; the 
frame where ${\bold p}=0$ for the $k$-integrals in (\ref{emq}) and (\ref
{emd}), while in the case of the
diquark current (\ref{dc}) one chooses the frame ${\bold k}=0$ to 
introduce the cut-off and then  
generalizes the result to an arbitrary frame. (This generalization is 
necessary because 
one has to insert the general Lorentz covariant 
parameterization of the diquark current into (\ref{emd}) before 
evaluating the $k$-integral.) 

The simplest case is the integral (\ref{emq}) for the scalar diquark: 
Besides scalar functions, the 
integrand involves terms $\propto k^{\nu}$ and $k^{\mu}k^{\nu}$, and 
these terms are most easily parameterized
in terms of Lorentz tensors made of $P^{\nu}\equiv (p+p')^{\nu}$ and $q^
{\nu}$. The most complicated
case is the integral (\ref{emd}) for the a.v. diquark: Here one has to 
insert the following parameterization
of the 3-point function (\ref{dc}) for the a.v. diquark \cite{QD1}:
\begin{eqnarray}
\Lambda_{\nu \lambda}^{\mu}(k',k) = F_1 (K,q)g_{\nu \lambda}K^{\mu} + 
F_2 (K,q)k'_{\nu} {g_\lambda}^{\mu}
+ F_3 (K,q)k_{\lambda} {g_\nu}^{\mu} \nonumber\\
+F_4 (K,q)k'_\nu k_\lambda K^{\mu}
+F_M(K,q) ({g_\lambda}^{\mu}q_\nu -{g_\nu}^{\mu}q_\lambda )], \label
{pard}
\end{eqnarray}
where $K=k'+k$ and $q=k'-k$, and the scalar functions depend on $K^2$, 
$q^2$ and $K \cdot q$. 
For $q\rightarrow 0$, $F_M$ is the ``magnetic moment'' of an off-shell 
a.v. diquark 
\footnote{For an on-shell a.v. diquark one naively expects \cite{QD1} 
$F_M(q=0)=2$ 
for the same reason why the
magnetic moment of the deuteron is naively expected to be the sum of 
the proton and neutron
magnetic moments. We confirmed that for an on-shell a.v. diquark this 
is indeed valid almost
independently of the choice of parameters. In our calculations to be 
discussed below, however,
the a.v. diquark is always unbound. In this case the ``magnetic 
moment'' depends on the virtuality
of the a.v. diquark and can assume values much less than 2.}. This 
parameterization is introduced
into (\ref{emd}), and a further parameterization is introduced in order 
to express 
$\lambda_D^{\mu}(p',p)$ in terms of $P^{\mu}$ and $q^{\mu}$.
In this way one derives the formal expressions for the diquark 
contributions to the 
Dirac-Pauli form factors of eq. (\ref{par}) in terms of Lorentz scalar
functions, and then one can set $q=0$ for the evaluation of the 
magnetic moments. 
Similarly, for the scalar-a.v. mixing terms one has the structure 
\begin{eqnarray}
\Lambda_{5 \lambda}^{\mu}(k',k) = F_{m}(K,q) \epsilon^{\mu}_{\lambda 
\alpha \beta} K^{\alpha} q^{\beta},
\label{parm}
\end{eqnarray}
which determines the mixing part of the vertex (\ref{emd}) and its 
contribution to the
Dirac-Pauli form factors of the nucleon.

\subsection{Axial and pion-nucleon coupling constants}

To obtain the isoscalar ($\alpha=0$) and isovector ($\alpha=3$) axial 
vector currents 
in the valence quark picture, we insert the operator $\tau^{\alpha} 
\gamma^{\mu} \gamma_5$ into
the diagrams of Fig.1, where $\tau^{0}\equiv 1$. The pion absorption 
current is obtained
similarly by inserting the operator $g \tau^{3} \gamma_5$, where the 
pion-
quark coupling constant $g$ is the residue of $\tau_{\pi}$ (eq.(\ref
{tau_s})) at the
pion pole. As we will explain at the end of this
subsection, in order to be consistent with the PCAC relation one has to 
take into account also
the contribution from the ``exchange diagram'' (Fig. 3) for the pion 
absorption current, 
even in the
static approximation used here. Therefore, separating again the isospin 
matrix elements, we write for the
axial vector coupling constants $g_A^{(\alpha)}$ and the pion-nucleon 
coupling constant $g_{\pi NN}$:
\begin{eqnarray}
g_A^{(0)} &=& {\cal G}_Q^s + {\cal G}_Q^a + 2{\cal G}_D^a  \label{pga0} 
\\   
g_A^{(3)} &=& {\cal G}_Q^s - \frac{1}{3} {\cal G}_Q^a + \frac{4}{3}
{\cal G}_D^a + \frac{2}{\sqrt{3}} {\cal G}_D^m 
\label{pga1} \\   
g_{\pi NN} &=& {\cal I}_Q^s - \frac{1}{3} {\cal I}_Q^a + \frac{4}{3}
{\cal I}_D^a + \frac{2}{\sqrt{3}} {\cal I}_D^m
- {\cal I}_E^s - \frac{5}{3} {\cal I}_E^a + \frac{1}{\sqrt{3}} {\cal I}
_E^m, \label{pgpi}
\end{eqnarray}
where the subscript $E$ in (\ref{pgpi}) refers to the exchange diagram 
of fig.3, 
and the meanings of the other 
symbols are as in the previous subsection. The quantities ${\cal G}$ 
and ${\cal I}$ introduced above are
obtained by expressing the axial and pion absorption currents for the 
quark diagram ($j_{AQ}^{\mu}$ and 
$j_{\pi Q}$), the diquark diagram ($j_{A D}^{\mu}$ and $j_{\pi D}$) and 
the exchange diagram ($j_{\pi E}$)
in terms of form factors as follows:
\begin{eqnarray}
j^{\mu}_{A R}(p',p) &=& {\bar \Gamma}_N^a (p') \lambda_{AR,ab}^{\mu}
(p',p) {\bar \Gamma}_N^b(p) \label{ja} \\
&\equiv& \sum_{d=s,a,m} {\bar u}_N(p') \left[{\cal G}_{R}^d(q^2) \gamma^
{\mu} \gamma_5 + {\cal H}_{R}^d(q^2) 
q^{\mu} \gamma_5 \right] u_N(p)  \label{ja1}  \\
j_{\pi R}(p',p) &=& {\bar \Gamma}_N^a (p') \lambda_{\pi R,ab}(p',p) 
{\bar \Gamma}_N^b(p) \label{jp} \\
&\equiv& \sum_{d=s,a,m} {\bar u}_N(p') \left[{\cal I}_R^d(q^2) \gamma_5 
\right] u_N(p)  \label{jp1},  
\end{eqnarray}
where $R=Q,D$ in eq. (\ref{ja1}) and $R=Q,D,E$ in (\ref{jp1}). 
The quantities ${\cal G}$ and ${\cal I}$
in eqs. (\ref{pga0})-(\ref{pgpi}) are then obtained by setting $q^2=0$ 
in the corresponding 
form factors defined in (\ref{ja1}) and (\ref{jp1}). 

The axial vertices $\lambda_{AQ}^{\mu}$ and $\lambda_{AD}^{\mu}$ are 
obtained from the corresponding expressions
(\ref{emq}) and (\ref{dc}) for the electromagnetic case by the 
replacement $\gamma^{\mu} \rightarrow \gamma^{\mu}\gamma_5$, 
while the pionic vertices $\lambda_{\pi Q}$ and $\lambda_{\pi D}$ are 
obtained by $\gamma^{\mu} \rightarrow g \gamma_5$.
The 3-point functions corresponding to (\ref{dc}), $\Lambda_{A}^{\mu}$ 
and $\Lambda_{\pi}$, 
have the following Lorentz structures:
\begin{eqnarray}
\Lambda_{A, \nu\lambda}^{\mu}(k',k)&=& {\epsilon^{\mu}}_
{\nu\lambda\sigma}
G(K,q)K^{\sigma} \label{ax1} \\
\Lambda_{A, 5 \lambda}^{\mu}(k',k) &=& 
{g^{\mu}}_{\lambda}G_{m1}(K,q) + K^{\mu}K_{\lambda}G_{m2}(K,q) \label
{ax2} \\
\Lambda_{\pi,\nu\lambda}(k',k) &=& 
{\epsilon^{\alpha\beta}}_{\nu\lambda} I (K,q)K_{\alpha}q_{\beta} \label
{pion1} \\
\Lambda_{\pi,5 \lambda}(k',k) &=& I_{m1}(K,q)K_{\lambda}+
I_{m2}(K,q)q_{\lambda}, \label{pion2} 
\end{eqnarray}
where for the axial vertices (\ref{ax1}), (\ref{ax2}) we show only the 
terms 
relevant in the limit $q\rightarrow 0$. 

Before giving the expression for the contribution of the exchange 
diagram ($\lambda_{\pi E}$) to the pionic
vertex (\ref{jp1}), we explain why it is necessary to consider this 
diagram even in the static approximation:
Let us first refer to the case of the electromagnetic interaction. The 
Ward-Takahashi identity for the
quark-photon vertex $\Gamma^{\mu}$ is
\begin{eqnarray}
S(k') q_{\mu}\Gamma^{\mu} S(k) = Q_q\left(S(k)-S(k')\right). \label{wt}
\end{eqnarray}  
Applying this identity to the quark-photon vertex in the exchange 
diagram (the cross in fig.3), 
we see that in the static
limit for the exchanged quark ($S(k)\rightarrow -1/M$), the r.h.s. of 
this
identity is zero. It is therefore consistent to assume that in the 
static approximation
there is no contribution from the exchange diagram, as has been done in 
previous works 
\cite{DQ}. \footnote{The same argument holds also for the baryon 
current, and therefore no
contributions from the exchange diagram are needed to satisfy the quark 
number sum rules
in the static approximation.} 
If there were no violations of gauge invariance due to the cut-off 
procedure, 
the electromagnetic current of the nucleon calculated 
from the quark and diquark diagrams alone would satisfy the current 
conservation $q_{\mu} J^{\mu}=0$.
Similarly, if we apply the axial Ward-Takahashi identity 
\begin{eqnarray}
S(k') q_{\mu}{\bold \Gamma}_A^{\mu} S(k) = {\bold \tau} \left(\gamma_5 S
(k) + S(k') \gamma_5\right)
+ 2 M S(k') \gamma_5 {\bold \tau} S(k) \label{awt}
\end{eqnarray}
to the quark axial vertex in the exchange diagram of fig.3, we note that 
in the static limit ($S(k)\rightarrow -1/M$) the r.h.s. of (\ref{awt}) vanishes, too, 
which suggests that the exchange
diagram should not be taken into account in the calculation of the 
axial vector current in the static approximation. If there were no 
violations of chiral invariance of different origin 
\footnote{In the present calculation, however, there are 2 origins of 
violation of PCAC and the Goldberger-
Treiman relation: The first reason is the omission of the vector 
diquark channel, as has been explained in sect. 2.  
The second, and probably more important, reason is that the 3-momentum 
(or Lepage-Brodsky)
cut-off scheme used here satisfies charge and baryon number 
conservation, 
but not the full vector current conservation
and PCAC. Since in the static approximation used in this work the quark-
diquark vertex function
is independent of the relative momentum, the dependence of the
quark-diquark loop integrals on the cut-off is enhanced considerably, 
and this leads to 
sizeable violations of the Goldberger-Treiman relation as will be 
discussed in sect. 5.},
the axial current of the nucleon
calculated from the quark and diquark diagrams alone would satisfy the PCAC 
relation (with the
pion pole contributions subtracted): $q_{\mu} J_A^{\mu}(q) = f_{\pi} J_
{\pi}(q)$, 
where $f_{\pi}=M/g$.
The pion absorption current $J_{\pi}$ in this relation, however, {\em 
includes} the effects 
of the exchange diagram, since in our above discussion the
last term in eq. (\ref{awt}) just corresponds to the contribution of the 
exchange diagram to $J_{\pi}$. 
In other words, if $g_A^{(3)}$
is calculated from the quark and diquark diagrams alone, the 
corresponding pion-nucleon 
coupling constant in the 
Goldberger-Treiman relation $M_N g_A^{(3)} = g_{\pi NN} f_{\pi}$ 
includes the effect of the exchange diagram.
In Appendix B we explicitly show the validity of the PCAC relation for 
the case where only the scalar
diquark channel is included, both in the exact Faddeev framework and in 
the static approximation.

The contribution of the exchange diagram to the pionic vertex of eq.
(\ref{jp}) in the static approximation
can be expressed in terms of the quark-diquark bubble graph $\Pi_N$ of 
eq. (\ref{pin}) by
\begin{eqnarray}
\lambda_{\pi E}^{ab} = -3 g \Pi_N^{ac}(p') \gamma_{c} \left(\frac{1}
{M^2} \gamma_5\right) 
\gamma_{d} \Pi_N^{db}(p), \label{gpex} 
\end{eqnarray}
which includes a color factor 3. If only the scalar diquark channel is 
included, the Faddeev kernel in the static 
approximation is
equal to $\Pi_N(p)\cdot 3/M$ (see (\ref{Faddeev})), which simplifies
the calculation of the exchange diagram contribution to the pion
absorption current (\ref{jp1}).  

The explicit evaluation of the various contributions shown in eqs. 
(\ref{pga0})-(\ref{pgpi}) in the covariant three-momentum cut-off scheme 
proceeds
along the lines discussed in the previous subsection.   

\subsection{Pion cloud effects}

Although the main purpose of this work is to extract information on the 
strength of the
a.v. diquark correlations in the nucleon, we also should estimate 
the size of the pion cloud effects in the present quark-diquark 
model.
The purpose of this subsection is to explain a simple way, involving 
various approximations,
to estimate these effects. These approximations, of course, should be 
avoided in a more
refined treatment of pion cloud effects.
(For a more general recent discussion on pionic effects on the static 
properties of the
nucleon, we refer to ref. \cite{LW}.) 

To estimate the role of pion cloud effects on the static properties of 
the nucleon,
we use an on-shell approximation for the ``parent quark'', which has 
been used also to estimate the pion cloud effects on 
the structure functions (see ref. \cite{conv}).
That is, the operator insertions on the quark lines
in fig. 1 are replaced by the diagrams of fig. 2, but these diagrams 
are evaluated by assuming
that the external quark lines are on the mass shell. For the 
calculation of $g_A^{(\alpha)}$ 
this simply means that we replace the 'bare' axial vector coupling 
constant of the
quark (which is 1) by $g_{Aq}^{(\alpha)}$, which includes the effect of 
the pion dressing, 
and similarly for $g_{\pi NN}$ we replace 
the bare $\pi qq$ coupling ($g$) by the dressed one ($g_{\pi qq}$) 
\footnote{In order to respect chiral
symmetry, one should also include the mixing between the pion and the 
sigma meson in the third diagram
of fig. 2, but this will not be done here in this rather schematic 
treatment of pion cloud 
effects. We also note that in principle the pion cloud effects lead to modifications
of the quark and diquark propagators and the quark-diquark vertex functions.
However, if the quark propagator is approximated by its
pole part, these modifications can be absorbed by a redefinition of the constituent
quark mass and the 4-fermi coupling constants, as has been discussed in ref. \cite{MIN}}. 
For the calculation of magnetic moments,
there appears in principle a new type of quark operator associated with 
the anomalous magnetic moment
of the constituent quark. In a  non-relativistic approximation, 
however, the matrix element of 
this operator can be related to the isoscalar and isovector axial vector
coupling constants of the nucleon, see eq. (\ref{kpi}) below. 

The magnetic moments of the $u$ and $d$ quarks including the pion 
dressing are written as
\begin{eqnarray}
\mu_u &=& Z_q\,Q_u + 3 {\cal F}_{\pi} \equiv Q_u + \kappa_u \label{muu} 
\\
\mu_d &=& Z_q\, Q_d + {\cal F}_q - {\cal F}_{\pi} \equiv Q_d + \kappa_d 
\label{mud}
\end{eqnarray}
The quantities ${\cal F}_{q}$ and ${\cal F}_{\pi}$ are obtained by 
expressing the results for the 
quark diagram $j^{\mu}_q$ (first diagram of fig. 2) and the pionic 
diagram
$j^{\mu}_{\pi}$ (second diagram of fig. 2) in terms of Dirac-Pauli form 
factors:
\begin{eqnarray}
j^{\mu}_r(p',p) &=& {\bar \Gamma}_q (p') \lambda_{r}^{\mu} (p',p) 
\Gamma_q(p) \label{jqq} \\
&\equiv&  {\bar u}_q(p') \left[{\cal F}_{1r}(q^2) \gamma^{\mu} + {\cal 
F}_{2r}(q^2) 
\frac{i \sigma^{\mu \nu} q_{\nu}}{2M} \right] u_q(p),  \label{parq} 
\nonumber
\end{eqnarray}
where $r=q, \pi$. The contributions to the magnetic moments in eq. (\ref
{muu}) and (\ref{mud})
are then obtained by
${\cal F}_q = {\cal F}_{1q}(0) + {\cal F}_{2q}(0)$ and
${\cal F}_{\pi} = {\cal F}_{1{\pi}}(0) + {\cal F}_{2{\pi}}(0)$.
The quark spinor $\Gamma_q(p)$ in (\ref{jqq}) is given by $\Gamma_q(p) 
= \sqrt{Z_q} u_q(p)$,
where the free Dirac spinor $u_q(p)$ involves the quark mass $M$, and 
the normalization factor, which also appears explicitly in (\ref{muu}) 
and (\ref{mud}), is given
by \cite{MIN}
\be
Z_q=\left( \left.1+\frac{\partial \Pi_q(k)}{\partial \fslash{k}}
\right|_{\fslash{k}=M} \right)^{-1}, \label{zq}
\ee
with the quark self energy 
\be
\Pi_q (k)=
3\int\frac{d^4 q}{(2\pi)^4}[\gamma_5 S_F(q) \gamma_5] \tilde{\tau}_{\pi}
(k-q). \label{pip}
\ee

The vertices $\lambda_{q}^{\mu}$ and $\lambda_{\pi}^{\mu}$ in (\ref
{jqq}) are obtained from the diagrams of
Fig.2 as follows:
\begin{eqnarray}
\lambda^{\mu}_{q}(p',p) &=& -\int \frac{d^4 k}{(2\pi)^4} S_F(k+q)\gamma^
{\mu}S_F(k)\tilde{\tau}_{\pi}(p-k)
\label{emqq} \\
\lambda^{\mu}_{\pi}(p',p)&=&-i \int \frac{d^4 k}{(2\pi)^4}S_F(p-k)\tau_
{\pi}(k')
\Lambda^{\mu}_{\pi}(k',k)\tau_{\pi}(k),
\label{emdq}
\end{eqnarray}
where the 3-point function for the electromagnetic current of the pion 
is given by
\begin{equation}
\Lambda^{\mu}_{\pi}(k',k)= 3i \int\frac{d^4 t}{(2\pi)^4 }
\mbox{tr}_D [S(t+q)\gamma^{\mu} S(t) \gamma_5 S(t-k)\gamma_5 ].
\label{dc1}
\end{equation}
The reduced pion t-matrix $\tilde{\tau}_{\pi}$ in the above expressions 
is defined so as to
exclude the contribution of the 'bare' contact interaction
\footnote{The contribution of the contact term to the vertex (\ref
{emqq}) is equivalent to
an ``RPA-type'' vertex correction induced by an interaction of vector 
type 
($\propto \left(\overline{\psi}\gamma^{\mu}\tau^{\alpha}\psi\right)^2$) with a 
strength determined
by the Fierz transformation of the pseudoscalar interaction term in the
lagrangian. 
However, our strategy is
to consider our underlying interaction Lagrangian in the $q\overline{q}
$ channels to be 
already Fierz symmetrized, and to discard all channels except the 
scalar-pseudoscalar one in order
not to introduce additional parameters (see sect. 2).}: 
$\tilde{\tau}_{\pi} \equiv {\tau}_{\pi} + 2iG_{\pi}$.   

Having determined the magnetic moments (\ref{muu}), (\ref{mud}) of the 
on-shell parent quark 
in the way described above, one should then use the quark operator
${\displaystyle \frac{i \sigma^{\mu \nu} q_{\nu}}{2M} \kappa_q}$, where 
${\displaystyle \kappa_q = \frac{1}{2}(1+\tau_3) \kappa_u + \frac{1}{2}(1-\tau_3) \kappa_d \equiv 
\kappa_q^{(0)} 
+ \tau_3 \kappa_q^{(3)}}$, to calculate the contribution to the nucleon 
magnetic moments
as described in the previous subsection. 
In a non-relativistic approximation for the constituent quark, however, 
this operator (for $\mu=i$) can be 
replaced by ${\displaystyle \frac{i}{2M} \left({\bold \gamma} \gamma_5 
\right) \times {\bold q}\,\kappa_q}$,
and then the contribution of the pion cloud to the anomalous magnetic 
moment of the nucleon can be written
as
\begin{eqnarray}
\kappa_{p(n)}^{(\pi)} = \frac{M_N}{M}\left(g_A^{(0)} \kappa_q^{(0)} \pm 
g_A^{(3)} \kappa_q^{(3)} \right)
\label{kpi}
\end{eqnarray}
with $+$ and $-$ for proton and neutron, respectively.

Similarly, the axial vector and $\pi qq$ coupling constants of the on-
shell parent quark 
including the pion dressing are written as
\begin{eqnarray}
g_{Aq}^{(0)} &=& Z_q + 3 {\cal G}_q \label{pga0q} \\   
g_{Aq}^{(3)} &=& Z_q - {\cal G}_q  \label{pga1q} \\   
g_{\pi qq} &=& Z_q\,g - {\cal I}_q.  \label{pgpiq}
\end{eqnarray}
In this case, there is no contribution due to the pion current (the 
third diagram in fig. 2).
These quantities ${\cal G}_q$ and ${\cal I}_q$ are
obtained by expressing the axial and pion absorption currents for the 
quark diagram ($j_{Aq}^{\mu}$ and 
$j_{\pi q}$) in terms of form factors as follows:
\begin{eqnarray}
j^{\mu}_{A q}(p',p) &=& {\bar \Gamma}_q (p') \lambda_{Aq}^{\mu}(p',p) 
{\bar \Gamma}_q(p) \label{jaq} \\
&\equiv& {\bar u}_q(p') \left[{\cal G}_{q}(q^2) \gamma^{\mu} \gamma_5 + 
{\cal H}_{q}(q^2) 
q^{\mu} \gamma_5 \right] u_q(p)  \label{ja1q}  \\
j_{\pi q}(p',p) &=& {\bar \Gamma}_q (p') \lambda_{\pi q}(p',p) {\bar 
\Gamma}_q (p) \label{jpq} \\
&\equiv& {\bar u}_q(p') \left[{\cal I}_q(q^2) \gamma_5 \right] u_q(p)  
\label{jp1q},  
\end{eqnarray}
The quantities ${\cal G}_q$ and ${\cal I}_q$
in eqs. (\ref{pga0q})-(\ref{pgpiq}) are then obtained by setting $q^2=0
$ in the form factors defined
in (\ref{ja1q}) and (\ref{jp1q}). 
The axial vertex $\lambda_{Aq}^{\mu}$ in (\ref{jaq}) is obtained from 
the corresponding expression
(\ref{emqq}) for the electromagnetic case by the replacement $\gamma^
{\mu} \rightarrow \gamma^{\mu}\gamma_5$, 
while the pionic vertex $\lambda_{\pi q}$ in (\ref{jpq}) is obtained by 
replacing $\gamma^{\mu} \rightarrow g \gamma_5$. The isoscalar and 
isovector axial vector 
coupling constants of the
nucleon are then renormalized simply by multiplying the factors $g_{Aq}^
{(0)}$ and 
$g_{Aq}^{(3)}$, respectively, while the $\pi NN$
coupling constant gets multiplied by $g_{\pi qq}/g$.    

\section{Results and discussions}

\setcounter{equation}{0}
In this section we will discuss our results for the nucleon structure 
functions and
static properties. First we have to explain the choice of our 
parameters. 
The three basic parameters of the NJL model, namely the 4-fermi 
coupling constant in the
pionic channel $G_{\pi}$, the UV cut-off $\Lambda$, and the current 
quark mass $m$, are determined
so as to reproduce the pion mass $m_{\pi}=140$ MeV as the pole of 
$\tau_{\pi}$ 
(eq.(\ref{tau_s})),  
the pion decay constant $f_{\pi}=93$ MeV via the familiar quark loop 
diagram for charged
pion decay \cite{NJL1}, and the constituent quark mass $M=400$ MeV via 
the gap equation. (The qualitative
behaviour of the results does not depend on this particular choice for 
$M$.) The resulting parameters
in the 3-momentum cut-off scheme are $G_{\pi}=6.92$ GeV$^{-2}$, 
$\Lambda=593$ MeV, 
and $m=5.96$ MeV.

In order to see the dependence of our results on the strength of the 
a.v. diquark
correlations, we consider the ratio $r_a=G_a/G_{\pi}$ as a free 
parameter, and 
adjust the strength in the scalar
diquark channel $r_s=G_s/G_{\pi}$ so as to reproduce the experimental 
nucleon mass $M_N=940$ MeV 
from the Faddeev equation in the static approximation (eq.(\ref
{Faddeev})).
Table 1 shows 3 particular choices for $r_a$. 
Case I refers to the pure scalar diquark model of ref. \cite{MIN}, 
which leads to a strongly bound 
scalar diquark of mass $M_s=596$ MeV.
For case III the value $r_a=0.66$ is strong enough to produce a bound 
delta state 
of mass $M_{\Delta}=1140$ MeV, but nevertheless the a.v. diquark is 
still unbound. 
The corresponding value of $r_s$ is much smaller than that in case I 
to give the same $M_N$. 
This case therefore refers to weak scalar diquark correlations with a 
weakly bound scalar diquark
of mass $M_s=$ 766 MeV. Case II ($r_a=0.25$)
describes a situation intermediate between these two.  
For each case we also show in Table 1 the corresponding ``weight'' 
$W_s$ 
of the scalar diquark channel, which is defined as the contribution to 
the 
baryon number (see eq.(\ref{norm})). 

In order to compare the calculated structure functions and quark LC 
momentum distributions
with the empirical ones, we have to associate a ``low energy scale'' 
$\mu^2\equiv Q_0^2$
to our NJL model results, and perform the $Q^2$ evolution up to the 
value of 
$\mu^2 = Q^2$ where experimental data and empirical parameterizations 
are available. 
For this purpose we solve the DGLAP equation \cite{DGLAP} in the 
$\overline{\rm MS}$ scheme
up to next-to-leading order,
using the computer code of ref.\cite{EVO} with $N_f=3$ and
$\Lambda_{QCD}=250$ MeV, and compare with empirical parameterizations 
which also refer
to the $\overline{\rm MS}$ scheme. If we determine $Q_0^2$ so as to 
reproduce
the overall features of the empirical valence quark distributions (see 
fig. 7 below), we obtain 
$Q_0^2=0.16$ GeV$^2$, i.e., $Q_0$ is equal to our constituent quark 
mass $M$.
  
\subsection{Structure functions}

Since the main purpose of this work is to investigate the role of a.v. 
diquark correlations
on the flavor dependence of the structure functions, we discuss this 
point first.
In Fig. 4 we compare the valence quark distributions (multiplied
by $x$), $\frac{1}{2}x\, u_v(x)$ and $x\,d_v(x)$, 
 without pion cloud effects,  
for the three cases I, II and III of Table 1. (The factor $\frac{1}{2}$ 
is multiplied in order to compare
distributions with the same normalization.) Case I shows a rather 
large 'flavor asymmetry', i.e., $d_v$ is
softer than $u_v$. This is a typical example for strong scalar diquark 
correlations, as has been discussed
first in ref. \cite{FLAV}: If $M_s$ is small, the scalar diquark 
carries a LC momentum fraction which is small
compared to the case of a weakly bound state ($M_s \simeq 2M$), and 
accordingly the 
spectator quark
carries a large fraction of the LC momentum. Since the scalar diquark 
consists of a $ud$ pair,
this implies that for strong scalar diquark correlations the $u$ quark 
distribution in the proton 
has a 'hard' component,
while the $d$ quark distribution has only a 'soft' component.  
In terms of the distributions given by (\ref{genu}) and (\ref{gend}), 
the hard component is
the term $F_{Q/P}^s(x)$ due to the quark diagram, while the soft 
component is $F_{Q(D)/P}^s(x)$ 
due to the diquark diagram. For case I, the two valence
$u$ quarks in the proton carry 71 $\%$ of the LC momentum, while the 
valence 
$d$ quark carries only 29 $\%$. This flavor asymmetry
is gradually decreased as we increase $r_a$, as is shown in Fig.4. 
For the cases II (III), the two valence $u$ quarks carry 69 $\%$ (68 $\%
$) of the total 
LC momentum. Particularly important is the fact that for large
$x$ the valence $d$ quark distribution increases rapidly with 
increasing $r_a$ and approaches
the valence $u$ quark distribution.

The influence of the a.v. diquark correlations on the flavor asymmetry 
of the valence quark distributions
is mainly an indirect one: If one increases $r_a$ one has to choose a 
smaller $r_s$ in order to get the
same nucleon mass, which implies that $M_s$ ($W_s$) increases 
(decreases) with increasing $r_a$. The direct
influence of the a.v. diquark channel on the flavour asymmetry is 
small, since even for 
case III the
a.v. diquark correlations are 'small' compared to the scalar ones in 
the sense that the a.v. 
diquark is still
unbound. In terms of the distributions (\ref{genu}) and (\ref{gend}), 
this means that $F_{Q(D)/P}^a(x) \simeq 2\,F_{Q/P}^a(x)$.
One should also note that, due to the same reason, the a.v. diquark 
correlations lead to a 
larger width of the valence
quark distributions: Since $M_s$ increases as we increase the ratio 
$r_a$, the binding energy 
of the diquark-quark system increases because the nucleon mass is 
fixed, and as a consequence the width of the valence quark distributions
increases.  

Let us now discuss the flavor dependence of our valence quark 
distributions in connection with the
empirical information. The quantity which is most sensitive to the 
flavor asymmetry of the
valence quark distributions is the ratio of the neutron to the proton 
structure function 
$R(x)=F_2^n(x)/F_2^p(x)$ for large $x$ \cite{MEL}. Flavor symmetric 
distributions 
($d_v(x)/u_v(x) = \frac{1}{2}$), like in the naive SU(6) quark model, 
give a constant value $R(x)=2/3$, while for the case of a strong 
dominance of scalar diquark correlations over the a.v. ones
one expects $d_v(x)/u_v(x) \rightarrow 0$, or $R\rightarrow 1/4$, as 
$x\rightarrow 1$. 
Our calculated ratios $R(x)$ for $Q^2=12$ GeV$^2$ are plotted for 
several values of $r_a$ 
in Fig. 5. The
influence of the a.v. diquark correlations on the flavor dependence of 
the valence quark
distributions is clearly seen: For large $x$ the ratio gradually 
increases with increasing
$r_a$ (decreasing strength of the scalar diquark correlations). To 
compare this behaviour with the
experimental data, one must note that $F_2^n$
has to be extracted from the experimental deuteron structure function
\cite{SLAC}, 
and therefore some model dependence 
is introduced. A very careful re-analysis has been performed in ref.\cite
{MEL}, including binding and off-shell effects.
The lower data points shown in Fig. 5 have been extracted from the SLAC 
proton and deuteron
data \cite{SLAC} by using a deuteron model with on-shell nucleons, while the upper 
data points are taken from the re-analysis of ref. \cite{MEL}. 
From Fig. 5 we see that the empirical data, in 
particular the
re-analyzed ones, seem to require some admixture of a.v. diquark 
components. The
comparison with the re-analyzed data favors values for $r_a$ which are not too far from case II 
($r_a=0.25$) of Table 1. 
Case III, which is characterized by strong a.v. diquark correlations 
with a bound delta state,  
is in clear contradiction with the data.
\footnote{As discussed in ref.\cite{MEL}, QCD calculations based on 
hard gluon exchange 
\cite{HG}
show that $d_v(x)/u_v(x) \rightarrow \frac{1}{5}$ ($R(x)\rightarrow 3/7
$) as $x\rightarrow 1$,
in agreement with quark counting rules \cite{QC}. 
In our NJL model calculation, the ratio $R(x)$ for large $x$ is 
determined mainly by
the strength of the a.v. diquark correlations.
One should note, however, that there is no fundamental relation 
between the a.v. diquark
correlations discussed here and the hard gluon exchange correlations 
discussed in \cite{HG}. 
For example, the $x\rightarrow 1$ behaviour of our distributions 
depends to some extent on the cut-off scheme
(although we did not investigate whether also the {\em ratio} is 
sensitive to the cut-off scheme),
and the
lines in Fig.5 simply reflect the change of the probability of the $S=1
$ quark pair without any
preference of $S_z=0$ pairs, in contrast to the hard gluon exchange 
model.} 
Later we will investigate whether values around $r_a\simeq 0.25$, 
corresponding 
to a dominant scalar diquark component with weight
$\simeq 93\,\%$, are favored also by other observables of the nucleon.

In fig. 6 we show the valence quark distributions for case II ($r_a=0.25
$) of Table 1, including 
pion cloud effects. The distributions calculated in the NJL model 
(solid lines) are
evolved up to $Q^2=4$ GeV$^2$ (dashed lines) and compared to  
the empirical 
distributions extracted in a recent analysis from the experimental data 
(dotted lines) 
\footnote{The pion cloud
effects taken into account in this work make the valence quark 
distributions softer, 
i.e., reduce their
peak heights and increase the support at low $x$, but do not give rise 
to any substantial 
flavor dependence.}.
Compared to the results of the pure scalar diquark model (case I of Table 1) 
presented in ref.\cite{MIN}, 
we note that the a.v.
diquark effects work towards a better agreement with the empirical 
distributions, mainly due to the
increase of the width of the distributions. 
 
The antiquark distributions shown in Fig. 7 for the case II are not 
sensitive
to the a.v. diquark correlations. As in the pure scalar diquark model 
of ref. \cite{MIN}, 
we see that the pion cloud effects
taken into account in this work do lead to an enhancement of $\overline
{d}$ over $\overline{u}$, 
but this
flavor asymmetry of the Dirac sea is too small for intermediate values 
of $x$. The calculated
value of the Gottfried sum $S_G=0.262$ agrees well with the 
experimental value 
($0.235 \pm 0.026$) \cite{EXP}, 
since at very low $x$ our calculated distributions show a rather large 
asymmetry. For the case II 
shown here,
$92\,\%$ of the total LC momentum is carried by the valence quarks at 
the low energy scale, and the
rest is carried by the sea quarks.    

Fig. 8 shows the proton structure function $F_{2p}(x)$ 
calculated for the case II 
at $Q^2=$ 4.5 GeV$^2$ and $Q^2=$ 15 GeV$^2$
including pion cloud effects, in comparison with the
experimental data of ref.\cite{NM}. It is seen that the present quark-
diquark model
calculation reproduces the overall features of the experimental 
structure function.

\subsection{Static properties}

We now turn to the dependence of the results for the static properties of the nucleon 
on the
a.v. diquark correlations.  The proton and neutron 
magnetic moments
are shown in Fig.9 as functions of the ratio $r_a$. In the pure scalar 
diquark
model ($r_a=0$), the magnetic moments are too small in magnitude, even if the 
pion cloud
effects are taken into account. The contributions due to the a.v. 
diquark channel
enhance the magnitudes of the magnetic moments, and for $r_a\simeq 0.2$ 
the experimental
values can reproduced. Since the neutron magnetic moment is more 
sensitive to 
$r_a$ than the proton one, the ratio ${\displaystyle |{\mu_p}/{\mu_n}|}
$ 
decreases with increasing $r_a$. Therefore, for large $r_a$, as would be
required by a bound delta state (case III of Table 1), not only the 
magnitude
of the magnetic moments but also their ratio becomes unreasonable.  

For the three values of $r_a$ corresponding to the cases I, II
and III of Table 1, these results are split into the various 
contributions in Table 2.
Most important is the quark diagram in the scalar channel (the term 
${\cal F}_Q^s$
in (\ref{mup}) and (\ref{mun})), which increases as one goes from case I
to case II, in spite of the reduced probability $W_s$ of the scalar diquark 
channel. For 
$\mu_p$, the diquark diagram in the a.v. channel (${\cal F}_D^a$) and 
the
mixing term (${\cal F}_D^m$) give further enhancements, in particular 
for large
values of $r_a$. The enhancement of $|\mu_n|$ is dominated by the quark 
diagram
in the a.v. channel (${\cal F}_Q^a$), which is absent for $\mu_p$, and 
therefore
the neutron magnetic moment is more sensitive to the a.v. diquark 
correlations. 
The pion cloud gives rise to the following anomalous magnetic moments 
of the constituent
quarks (see eq. (\ref{muu}), (\ref{mud})): $\kappa_u=0.26$, $\kappa_d=-
0.29$. 
Due to eq. (\ref{kpi}), the pionic contributions to the nucleon magnetic 
moments are 
mainly of isovector type, too, and enhance their magnitudes, as is well 
known
\cite{LW}.    

The results for the axial vector coupling constants are shown in Fig. 
10 as
functions of the ratio $r_a$. The fact that the calculated $g_A^{(3)}$ 
is too small 
compared to the experimental value is most probably due to the static 
approximation
to the Faddeev equation used in the present calculation
\footnote{In the exact Faddeev calculation \cite{ISH}, the quark 
diagram in the scalar channel 
(${\cal G}_Q^s$ in eq. (\ref{pga1})) gives a larger contribution, and 
also the 
exchange diagram gives some enhancement. In this connection one should 
also note that the
charge radii of the nucleons are too small in the static approximation 
\cite{DQ}, 
indicating that relativistic effects are enhanced. This is one of the reasons
why the axial vector coupling constants are smaller than in the exact 
Faddeev calculation.}.
The a.v. diquark channel gives rise to a rather slow increase of $g_A^
{(3)}$ and a rapid
decrease of $g_A^{(0)}$. Since the increase of $g_A^{(3)}$ is slow and 
there is no
sizeable enhancement beyond $r_a\simeq 0.3$, one may conclude that a.v. 
diquark
correlations work in the desired direction, but there is no necessity 
to introduce
strong correlations as would be required by a bound delta state (case 
III of Table 1).
Similarly, the largest part of the reduction of $g_A^{(0)}$ occurs for 
small $r_a$,
whereas for larger $r_a$ the curve saturates.

Table 2 shows the various contributions to the axial coupling constants 
for the cases
I, II and III.
The most sizeable effects due to the a.v. diquark channel are the positive 
mixing term (${\cal G}_D^m$ in eq.(\ref{pga1})) for $g_A^{(3)}$, and 
the reduction of the 
scalar channel contribution (${\cal G}_Q^s$). The net effect is the 
rather slow increase
of $g_A^{(3)}$ and the rapid decrease of $g_A^{(0)}$ discussed above. 
The pion cloud renormalizes the axial vector coupling constants of the 
constituent quark
(see (\ref{pga0q}) and (\ref{pga1q})): $g_{Aq}^{(3)}=0.86$, $g_{Aq}^
{(0)}=0.78$.
These reductions due to the pion cloud are well known, see e.g., ref.
\cite{SUZ}. There it was pointed out that, due to the p-wave 
coupling of the
pion to the quarks, some amount of the nucleon spin carried by the 
quark spin is 
transfered to orbital angular momentum of the pion cloud, that is, to 
orbital angular
momentum of the sea quarks, leading to the reduction of $g_{Aq}^{(0)}$. 

In Fig. 11a we show the dependence of $g_{\pi NN}$ on the ratio $r_a$. 
In this case,
again, the result of the pure scalar diquark model is too small in 
comparison with
the experimental value, and the a.v. diquark channel gives rise to a 
sizeable
enhancement, in particular in the region of small $r_a$ where the 
experimental
value can be reproduced for $r_a\simeq 0.3$. Similarly to the cases of 
$g_A^{(3)}$
and $\mu_p$, the behaviour of the curves in Fig. 11a indicates that most 
of the a.v. diquark
correlation effects are exhausted already for 
admixtures which
are small compared to the case III of Table 1 (that is, the case of a bound delta state). 
Pion cloud effects renormalize the $\pi qq$ coupling constant
(see eq. (\ref{pgpiq})) from the bare value $g=4.23$ to $g_{\pi qq}=3.48$.  
Among the various contributions shown in Table 2, the quark diagram in 
the scalar
channel gives the dominant contribution, followed by the diquark 
diagram in the
a.v. channel and the exchange diagrams. There are large cancellations 
among the
exchange diagram contributions, leaving a net positive correction.  

Compared to the case of $g_{A}^{(3)}$, the $\pi NN$ coupling constant 
shows a rather
rapid increase in the small $r_a$ region, which indicates a violation 
of the Goldberger-Treiman
(GT) relation. In Fig. 11b we plot the ratio 
$\Delta_{GT}\equiv f_{\pi}\, g_{\pi NN}/M_N\, g_A^{(3)}$ as a function 
of $r_a$ 
(upper pair of lines). This figure shows that already for the pure 
scalar diquark
model there is a sizeable violation of the GT relation of about $13 
\%$ (for the
case including the pion cloud effects), and if the a.v. diquark channel 
is included
this violation increases up to $90\%$. This fact  
indicates a major problem of the static approximation to the Faddeev 
equation:
In this approximation the quark-diquark vertex function is independent 
of the
relative momentum in the quark-diquark system, and therefore all 
contributions
to the loop integrals which are not sufficiently damped due to the 
internal quark or diquark
propagators are very sensitive to the cut-off \footnote{As was 
discussed in the previous secions, 
the three-momentum cut-off procedure
used here violates the GT relation, although it conserves baryon 
number, electric charge
and momentum.}. The terms with the highest degree of divergence are 
those due to the
contact terms $4iG_s$ and $4iG_a g^{\mu \nu}$ in the scalar and a.v. t-
matrices 
in (\ref{tau_s}) and (\ref{tau_a}), respectively. 
In order to demonstrate that these terms are largely responsible for the
violation of PCAC, we show by the lower pair of lines
in Fig. 11b the result obtained by leaving out these contact terms. 
(Since the
contact terms act attractively, we have to choose larger values of 
$r_s$ for each 
$r_a$ in order to keep the nucleon mass fixed.) In such a calculation, 
where the
pole term due to the scalar t-matrix gives the dominant contribution 
besides small
corrections due to the $qq$ continuum terms, the violations 
of the GT relation are drastically reduced. This indicates that the large
violations of the GT relation will disappear as soon as one uses the
exact Faddeev wave functions. 
 
The main purpose of this subsection was to see whether our conclusions 
on the strength of the a.v. diquark correlations,
derived in sect. 5.1 from the flavor dependence of the quark LC momentum 
distributions,
are consistent with the static properties of the nucleon or not. From 
our
above discussions we can now conclude that the static properties, too, 
indicate
the necessity of some amount of a.v. diquark correlations, but not 
strong ones.
The range $0.15<r_a <0.3$, corresponding to $0.98>W_s>0.90$, seems reasonable 
from both points of view.

\section{Summary and conclusions}

\setcounter{equation}{0}
The purpose of this paper was to extract information on the strength
of the quark-quark interaction in the axial vector (a.v.) diquark 
channel
by comparing the results for the structure functions and 
the static properties of the nucleon with the empirical information.
For this purpose we used the Nambu-Jona-Lasinio (NJL) model in the 
framework
of a simple quark-diquark approximation (``static approximation'') to 
the full Faddeev equation.
The effects of the pion cloud were taken into account by assuming
an on-shell approximation for the parent quark, which in the case
of the quark distributions leads to the usual one-dimensional 
convolution
formalism. Our results are summarized as follows:

First, the observed flavor dependence of the structure functions implies
that the interaction in the a.v. diquark channel should be relatively
weak compared to that in the scalar diquark channel. In the model
used here we found that, in order to reproduce the flavor dependence,
the weight of the a.v. diquark component in 
the nucleon state (defined here as the contribution to the baryon
number) should not be much larger than  $10\%$.  Since in this case
a large part of the binding energy has to come from the correlations in 
the
scalar diquark channel, this implies rather strong scalar diquark 
correlations
in the nucleon. 

Second, the static properties of the nucleon indicate that some amount 
of a.v.
diquark correlations are required, in particular for the magnetic
moments which are too small in magnitude in a pure scalar diquark model.
The a.v. diquark channel gives
beneficial contributions to all static properties considered here, 
but it is neither necessary nor preferable to introduce strong a.v.
correlations. That is, a large part of the effects due to the a.v. 
channel on the
static properties is exhausted already for an admixture of less than 
$10\%$.   

Combining these two observations, we conclude that an admixture of the 
a.v. channel between $2\,\%$ and $10\,\%$
seems very reasonable. In terms of the 4-fermi coupling constants,  
this corresponds to $r_s\simeq 0.63$ and $r_a \simeq 0.25$ (see case II
of Table 1). It is
interesting to note that these values are similar to $r_s = 0.5$ and $r_a = 0.25$, 
which correspond to the 'color current' type interaction lagrangian \cite
{STAT}.
This relatively small value of $r_a$ 
implies rather weak a.v. diquark correlations,
since for example the a.v. diquark is unbound and no bound state for the
delta isobar can be obtained. While this latter fact at first sight 
seems to
indicate a problem of the model, we should note that for reasonable 
values of the
constituent quark mass the delta isobar always emerges very close to 
the (unphysical)
3-quark threshold in a model without confinement. (A similar situation 
holds also
for the vector mesons if they are described as $q\overline{q}$ bound 
states.)
Our results support the viewpoint that one should {\em not} attempt to 
describe
these heavier hadrons as bound states in a model without confinement. 
To describe
these states, one should at least incorporate one important aspect of 
the
confinement, namely the absence of unphysical thresholds for the decay 
into
colored states \cite{QD,THRES,INF} \footnote{It is interesting to note that 
recently a similar
conclusion has been derived for the description of nuclear matter \cite
{NEW}: In the
NJL model without confinement effects, a nucleon in the medium becomes a
loosely bound state near the threshold, and the resulting equation of 
state
does not saturate. If the thresholds are avoided, for example by introducing
a proper time infra-red cut-off \cite{INF}, the equation of state saturates.}.

The numerical results obtained for an admixture between $2\,\%$ and $10\,\%$  of the 
a.v. channel
show that the overall picture which emerges for the structure functions 
and
the static properties is quite reasonable. There are, however, several points 
which
should be improved, in particular concerning the violation of the 
Goldberger-Treiman
relation, the shape of $d_v(x)$, 
the difference $\overline{d}(x)-\overline{u}(x)$, etc. For these purposes, 
further work should be done towards a
full Faddeev description and a more refined treatment of pion cloud 
effects.        

\vspace{1 cm}

{\sc Acknowledgements} 

This work was supported by the Grant in Aid for Scientific Research of 
the
Japanese Ministry of Education, Culture, Sports, Science, and Technology, 
Project No. 
C2-13640298, and Project No. C2-11640257. The authors thank M. Miyama and 
S. Kumano for providing them with
the computer code of ref.\cite{EVO} for the $Q^2$ evolution, W. 
Melnitchouk for
the data points shown in Fig. 5 (Fig. 3 of ref.\cite{MEL}), and A.W. 
Thomas for
helpful discussions on structure functions and pion cloud effects.

\newpage

\section*{Figure Captions}

\begin{enumerate}

\item Graphical representation of the quark LC momentum distribution in the nucleon. 
The single (double) line
denotes the constituent quark propagator (diquark t-matrix), the hatched circle the 
quark-diquark vertex
function, and the operator insertion stands for $\gamma^+ \delta(k_--p_-
x)(1\pm \tau_z)/2$ for 
the U(D) quark distribution. 

\item Diagrams representing the dressing of the constituent quark by the 
pion cloud. The dashed line denotes the $q \overline{q}$
t-matrix in the pionic channel.

\item Feynman diagram where the operator insertion is made on the quark 
exchanged between the diquark and
the spectator quark. For the quark LC momentum distributions and the 
electromagnetic current, this
diagram does not contribute in the static approximation. However, it 
should be considered for the
calculation of $g_{\pi NN}$ as explained in the text.

\item Comparison of the valence quark distributions $x\,u_v(x)/2$ and 
$x\,d_v(x)$ for the 
cases I (dashed
lines), II (solid lines), and III (dotted lines) of Table 1.

\item Ratio of neutron to proton structure functions for several values 
of the ratio
$r_a$. For each $r_a$, the value of $r_s$ is determined so as to reproduce the
experimental nucleon mass. The data
points, which are taken from Fig. 3 of ref. \cite{MEL}, are based on 
the SLAC proton and 
deuteron data
using a deuteron model with on-shell nucleons (upper data points), and including 
binding and off-shell effects (lower data points).    

\item Valence quark distributions (multiplied by $x$) corresponding to 
case II of Table 1. 
The solid lines show the
input distributions ($\mu^2=Q_0^2$) calculated in the NJL model, the 
dashed lines show the
distributions obtained by the QCD evolution up to $Q^2=4$ GeV$^2$ in 
next-to-leading order, and the
dotted lines are the empirical distributions of ref.\cite{PAR}.

\item Same as Fig. 6 for the sea quark distributions.

\item Results for the proton structure function $F_2^p(x)$ at $Q^2=$ 4.5 
GeV$^2$ (a) and 
$Q^2=$ 15 GeV$^2$ (b) for case II of
Table 1, in comparison with the experimental data of ref.\cite{NM}.

\item The magnetic moment of the proton (a) and the neutron (b) as 
functions of $r_a$. 
The dashed lines are calculated in the valence quark picture, while the 
solid lines
include the effects of the pion cloud as described in the text. For each
$r_a$, the value of $r_s$ is determined so as to reproduce the 
experimental nucleon mass.  

\item Same as Fig. 9 for the isovector (a) and the isoscalar (b) axial 
vector coupling
constants of the nucleon.

\item Same as Fig. 9 for the $\pi NN$ coupling constant (a) and the 
ratio $\Delta_{GT}=
f_{\pi} g_{\pi NN}/M_N g_A^{(3)}$ (b). For $\Delta_{GT}$ we also show 
the result obtained
by subtracting the contact terms of the t-matrices in the 
scalar and a.v. diquark channels, as explained in the text. 

\item The axial vector vertex of the constituent quark including the 
pion pole terms.
The bubbles graphs here involve $q \overline{q}$ states in the pionic 
channel.
 
\end{enumerate}

\newpage
\appendix
{\LARGE Appendices}
\section{Diagonalization of the Faddeev kernel in the static 
approximation}

\setcounter{equation}{0}
In this Appendix we directly diagonalize the Faddeev kernel in the 
static approximation
(see eq.(\ref{Faddeev}) for the $T=\frac{1}{2}$ states) and derive the 
form of the nucleon vertex 
function as given in eqs. (\ref{wf1s})-(\ref{wf2}). 
The diagonalization will be done in the rest frame of the nucleon ($p^
{\mu}=(M_N,{\bold 0})$),
and the solution will be boosted to a general frame.

In order to remove the Dirac $\gamma$ matrices from the quark exchange 
kernel $Z$ 
(eq.(\ref{exk})), we first apply the following unitary transformation: 
\begin{eqnarray}
\tilde{K}(p) = \Omega K(p) \Omega^{\dagger} \equiv \tilde{Z} \, \tilde
{\Pi}_N(p)  \label{u1}
\end{eqnarray} 
where $\Omega={\rm diag}\left(\gamma_5, \gamma^{\mu}\right)$ and $\tilde
{Z}=\Omega Z \Omega^{\dagger}$,
$\tilde{\Pi}_N(p)=\Omega \Pi_N(p) \Omega^{\dagger}$. $\tilde{Z}$ has 
the form
\be
\tilde{Z} = \frac{3}{M}\left(
\begin{array}{ccc}
1 & -\sqrt{3} & -\sqrt{3} x^T \\
-\sqrt{3} & -1 & x^T \\
-\sqrt{3} x & x & X 
\end{array}
\right), \,\,\,\,\,\,\,{\rm where} \,\,\, 
X = \left(
\begin{array}{ccc}
-1 & 1 & 1 \\
1 & -1 & 1 \\
1 & 1 & -1 
\end{array}
\right) \label{ztilde}
\ee
and $x^T=(1,1,1)$. In the rest frame of the nucleon, the transformed 
quark-diquark bubble 
graph $\tilde{\Pi}_N(p)$ involves only the Dirac matrices $1$ and 
$\gamma_0$, and can be
written as follows:
\begin{eqnarray}
\tilde{\Pi}_N(p) &=& \tilde{G}_+(p) \frac{1-\gamma_0}{2} + \tilde{G}_-
(p) \frac{1+\gamma_0}{2}
\label{bubb1} \\
&\equiv& \tilde{\Pi}_{N(+)}(p) + \tilde{\Pi}_{N(-)}(p) \label{bubb2}  
\end{eqnarray}
with   
\be
G_+(p) = \left(
\begin{array}{ccc}
a & 0 & 0 \\
0 & b & c\,x^T \\
0 & c\,x & d\,I 
\end{array}
\right), \,\,\,\,\,\,\,
G_-(p) = \left(
\right), \label{gpm}
\ee
where $I = {\rm diag}\left(1,1,1\right)$ and 
\begin{eqnarray}
a(p)&=&\int \frac{d^4 q}{(2\pi)^4 } \frac{(p_0 -q_0 )
+M}{(p-q)^2 -M^2}\, \tau_s (q)\label{a}\\
b(p)&=& \int \frac{d^4 q}{(2\pi)^4 } \frac{-(p_0 -q_0)+M}
{(p-q)^2 -M^2} \,\tau_a^{00}(q)\label{b}\\
c(p)&=& \int \frac{d^4 q}{(2\pi)^4 } \frac{q^i}{(p-q)^2 -M^2}\, \tau_a^
{0i}(q)
\,\,\,\,\,\,\,({\rm no\,\,sum})
\label{c}\\
d(p) &=& \int \frac{d^4 q}{(2\pi)^4 }
\frac{(p_0 -q_0 ) +M}{(p-q)^2 -M^2}\, \tau_a^{ii}(q)
\,\,\,\,\,\,\,({\rm no\,\,sum}). 
\label{d}
\end{eqnarray}
In eqs. (\ref{c}) and (\ref{d}) the index $i$ is fixed to be any among 
$i=1,2,3$, i.e.;
one can replace $q^i\, q^i \rightarrow \bold q^2/3$ in these relations.
The quantity $\tilde{\Pi}_{N(+)}(p)$ in (\ref{bubb2}) gives the 
positive parity part of the
transformed kernel, while $\tilde{\Pi}_{N(-)}(p)$ refers to the 
negative parity
In the following,
we will refer only to the positive parity part without changing the 
notation, i.e, 
$\tilde{\Pi}_N \equiv \tilde{\Pi}_{N(+)}$,  
$\tilde{K} \equiv \tilde{Z} \tilde{\Pi}_{N(+)}(p)$, etc.
\footnote{The reason is as follows: In the rest frame of the nucleon 
the projection operator
onto positive parity is (in the original representation) 
$P_+ = {\rm diag}\left(p_+, p_-, p_+, p_+,p_+\right)$, where
$p_{\pm} = \left(1 \pm \gamma_0\right)/2$, and
the diagonal matrix refers to the 5 diquark indices $a=5$ (scalar 
diquark), 
$a=0$ (time component of the a.v. diquark) and $a=1,2,3$ (space 
components of the a.v. diquark).
After applying the unitary transformation $\Omega={\rm diag}\left
(\gamma_5, \gamma^{\mu}\right)$ as
in (\ref{u1}), this becomes $\tilde{P}_+ = {\rm diag}\left(p_-, p_-, p_-
, p_-,p_-\right)$.}. 

To the transformed eigenvalue equation $\tilde{\Gamma}(p) = \tilde{K}
(p) \tilde{\Gamma}(p)$, we
apply a further orthogonal transformation in order to bring the kernel 
into a block form corresponding
to $J=\frac{1}{2}$ and $J=\frac{3}{2}$ states:
\be
\hat{K}(p) = U \tilde{K}(p) U^t \equiv \hat{Z}\,\hat{\Pi}_N(p) \label
{u2}
\ee
with $U={\rm diag}\left(1,1,Y\right)$ and the $3\times 3$ matrix $Y$ 
diagonalizes $X$ of eq.(\ref{ztilde}):
\be
Y =
\left(
\begin{array}{ccc}
1/\sqrt3 & 0  & -2/\sqrt6 \\
1/\sqrt3 & -1/\sqrt2 & 1/\sqrt6 \\
1/\sqrt3 & 1/\sqrt2 & \sqrt6/3
\end{array}
\right). \label{y}
\ee
The transformed quantities $\hat{Z}$ and $\hat{\Pi}_{N}(p)$ are given 
by the block forms
\be
\hat{Z} = \frac{3}{M}\left(
\begin{array}{ccccc}
1 & -\sqrt{3} & -3 & & \\
-\sqrt{3} & -1 & \sqrt{3} & & \\
-3 & \sqrt{3} & 1 & &  \\
 & & & -2 &  \\
 & & & & -2 
\end{array}
\right), \label{zh}
\ee
\be
\hat{\Pi}_{N}(p) = \left(
\begin{array}{ccccc}
a & & & &   \\
 & b & \sqrt{3}\,c & & \\
 & \sqrt{3}\,c & d & & \\
 & & & d & \\
 & & & & d 
\end{array}
\right)\,\,\frac{1-\gamma_0}{2}, \label{hat} 
\ee
and the eigenvalue equation $\hat{\Gamma}(p)=\hat{Z} \hat{\Pi}_{N}(p)
\hat{\Gamma}(p)$ 
separates into two simple equations, one with dimension $3\times 2=6$ 
and one with $2\times 2=4$.
The latter one clearly corresponds to $J=\frac{3}{2}$, and the former 
one to $J=\frac{1}{2}$,
since there are 3 basis states corresponding to (i) the coupling of a 
scalar diquark and a quark,
(ii) the time component of the a.v. diquark and a quark, and (iii) the 
space components of the a.v.
diquark and a quark, and each of these three states has two spin 
directions.  

The solutions for the nucleon vertex function in this representation 
therefore have the
form
\begin{eqnarray}
\hat{\Pi}_{N\uparrow} = \left(
\begin{array}{c}
\alpha_1 \\ \alpha_2 \\ \alpha_3 \\ 0 \\ 0 
\end{array}
\right)_{\rm diquark} \bigotimes \left(
\begin{array}{c}
0 \\ 0 \\ 1 \\ 0 
\end{array}
\right)_{\rm quark}, \,\,\,\,\,
\hat{\Pi}_{N \downarrow} = \left(
\begin{array}{c}
\alpha_1 \\ \alpha_2 \\ \alpha_3 \\ 0 \\ 0 
\end{array}
\right)_{\rm diquark} \bigotimes \left(
\begin{array}{c}
0 \\ 0 \\ 0 \\ 1 
\end{array}
\right)_{\rm quark}, \label{sol}
\end{eqnarray}
corresponding to the two spin projections, where $(\alpha_1, \alpha_2, 
\alpha_3)$ is the
eigenvector of the upper $3\times 3$ block of $\hat{K}$ in (\ref{u2}) 
with the largest 
eigenvalue $\lambda_N(p)$.
The nucleon mass is then determined by $\lambda_N(p_0=M_N) = 1$.  

It is then easy to transform the solutions (\ref{sol}) back to the 
original representation
using $\Pi_N(p) = \Gamma^{\dagger} U \,\hat{\Pi}_N(p)$, and then to 
apply a boost which is the
product of a Lorentz transformation for spinors and ordinary 4-vectors, 
respectively, both with
velocity ${\bold v}=-{\bold p}/E_N(p)$ with $E_N(p)=\sqrt{M_N^2+{\bold 
p}^2}$. 
The result is given by eqs. (\ref{wf1s}), (\ref{wf1v}), where
\begin{eqnarray}
{\displaystyle \epsilon_{\lambda}^{\mu}(p) = \left(
\frac{{\bold p}\cdot {\bold \epsilon}_{\lambda}}{M_N}, \,\,\,
{\bold \epsilon}_{\lambda} + \frac{{\bold p}\, \left({\bold p}\cdot 
{\bold \epsilon}_{\lambda}\right)}
{M_N(E_N(p)+M_N)} \right)}, 
\label{eps}
\end{eqnarray}
where ${\bold \epsilon}_{\lambda}$ are the usual spherical unit vectors.

To derive the covariant form (\ref{wf2}) from (\ref{wf1v}), one makes 
use of the relations
\begin{eqnarray}
\left(1 \frac{1}{2}, \lambda s'|\frac{1}{2} s\right) &=& \frac{(-1)^
{\lambda+1}}{\sqrt{3}}
\left(\sigma^{[1]}_{-\lambda}\right)_{s's} \label{we} \\
\sum_{s'} \left({\bold \sigma}\cdot {\bold p}\right)_{s' s} \chi_{s'} 
&=&
\left({\bold \sigma}\cdot {\bold p}\right) \chi_s \label{spin},
\end{eqnarray}    
where (\ref{we}) follows from the Wigner-Eckert theorem, and $\chi$ in 
(\ref{spin}) is a 2-component
Pauli spinor. 

The diagonalization of the $T=\frac{3}{2}$ kernel of eq.(\ref{kd}) 
proceeds in the same way by applying
the unitary transformations $\Gamma$ and $U$ as above. The positive 
parity part of the kernel then
separates into a $2\times 2$ block for $J=\frac{1}{2}$ (corresponding 
to the coupling of the time
component and the space components of the a.v. diquark with the quark, 
respectively) and a diagonal
$2 \times 2$ block for $J=\frac{3}{2}$. The latter one has the same 
form as the 
lower $2\times 2$ block of the kernel (\ref{u2}), but with the factor $-
2$ in (\ref{zh}) 
replaced by $4$. The delta mass is therefore determined
by the equation ${\displaystyle \frac{12}{M} \, d(p_0=M_{\Delta})=1}$, 
and after 
transforming back to the original
representation and boosting the vertex function has the form of a 
Rarita-Schwinger spinor:
\begin{eqnarray}
\Gamma_{\Delta}^{\mu}(p) = b\,\, \sum_{\lambda s'} \left(1 \frac{1}{2}, 
\lambda s'|\frac{3}{2}\right)
\epsilon_{\lambda}^{\mu}(p)\, u_N(p,s'), \label{rss}
\end{eqnarray}
where $b$ is a normalization constant. 

\newpage

\section{PCAC in the Faddeev framework}

\setcounter{equation}{0}
In this Appendix we demonstrate the validity of the PCAC relation for 
the case of the scalar
diquark channel only, both in the exact Faddeev framework and in the 
static approximation.
For more detailed discussions we refer to ref. \cite{ISH}.

In the exact Faddeev framework, the isovector axial vector current of 
the nucleon 
is given by the sum of the
'quark diagram' and the 'exchange diagram' (see fig.3) as follows:
\begin{eqnarray}
j_A^{\mu}(q) &=& \int \frac{{\rm d}^4 k}{(2\pi)^4} \int \frac{{\rm d}^4 
k'}{(2\pi)^4}
\overline{\Gamma}_{N,P'}(k') \nonumber \\
&\times& \left[ S_F(\frac{P}{2}+k') \Gamma_q^{\mu} S_F(\frac{P}{2}+k) 
\tau_s(\frac{P}{2}-k) \delta(k'-k-\frac{q}{2})\right.  \label{qq} \\ 
&+& \left. 3 S_F(\frac{P'}{2}+k') \gamma_5 F_F(k+k'+\frac{q}{2}) 
\Gamma_q^{\mu} S_F(k+k'-\frac{q}{2})
\gamma_5 S_F(\frac{P}{2}+k)\right. \nonumber \\ 
&\times& \left. \tau_s(\frac{P'}{2}-k') \tau_s(\frac{P}{2}-k) \right] 
\Gamma_{N,P}(k) \label{ec}
\end{eqnarray}
Here the assignments of the momenta are as in refs. \cite{FAD2,ASA}, 
and the matrices 
$C\,\tau_2$, which appear in the
2-body vertex functions in the exchange diagram, have been processed as 
explained in 
ref. \cite{ASA}. The Faddeev equation for the 
nucleon spinor reads
\begin{eqnarray}
\Gamma_{N,P}(p) =  \int \frac{{\rm d}^4 p'} Z(p,p') S_F(\frac{P}{2}+p') 
\tau_s(\frac{P}{2}-p') \Gamma_{N,P}(p')
\label{fe}
\end{eqnarray}
with the quark exchange kernel $Z(p,p')=-3 \gamma_5 S_F(p'+p) \gamma_5
$. The quark axial 
vector vertex $\Gamma_q^{\mu}$
including the vertex corrections describing the pion pole is given by 
(see fig. 12)
\begin{eqnarray}
\Gamma_q^{\mu}(q) = \gamma^{\mu} \gamma_5 \tau_3 -i \Pi^{\mu}(q) \tau_3 
\gamma_5 \tau_{\pi}(q), 
\label{vert}
\end{eqnarray}
where $\Pi^{\mu}(q)$ is the bubble graph describing the pion-axial 
vector mixing:
\begin{eqnarray}
\Pi^{\mu}(q) = 12 i M q^{\mu} \int \frac{{\rm d}^4 k}{(2\pi)^4} 
\frac{1}{\left(k^2-M^2+i\epsilon\right)
\left((k+q)^2-M^2+i\epsilon\right)} \label{expl}
\end{eqnarray}
Using the definition of $f_{\pi}$, and defining the 'physical' pion 
propagator by 
${\displaystyle \Delta_{\pi}(q) \equiv \frac{-i}{g^2} \tau_{\pi}}$ so 
that its
pole term is simply given by ${\displaystyle \frac{1}{q^2-m_{\pi}^2}}$ 
with unit residue,
it is easy to show that the vertex (\ref{vert})
satisfies the axial Ward-Takahashi identity of the linear sigma model:
\begin{eqnarray}
q_{\mu} \Gamma_q^{\mu}(q) = \tau_3 \left(S^{-1}(p')\gamma_5 + \gamma_5 
S^{-1}(p) \right) - 
2 c \Delta_{\pi}(q) \Gamma_{q \pi} \label{wtq}
\end{eqnarray}
with $\Gamma_{q \pi} = \tau_3 \gamma_5 g$ and $c=f_{\pi} m_{\pi}^2$. 
We also note that if the pion
pole terms are removed from the beginning, as is sufficient for the 
calculation of $g_A$ (i.e., if only the first 
term in fig. 12 is taken for the quark
axial vector vertex), the corresponding axial Ward-Takahashi identity 
is obtained from (\ref{wtq}) by the
replacement $- c \Delta_{\pi}(q) \rightarrow \frac{M}{g}$, which is 
equal to $f_{\pi}$.  
Using (\ref{wtq}) in (\ref{ec}) and using the Faddeev equation (\ref
{fe}) together with the corresponding
equation for $\overline{\Gamma}_{N,p'}$, it is easy to derive the PCAC 
relation
\begin{eqnarray}
q_{\mu} j_A^{\mu}(q) = - 2 c \Delta_{\pi}(q) j_{\pi}(q) \label{pcac}
\end{eqnarray}
where the pion absorption current is given by (\ref{ec}) with the 
replacement 
$\Gamma_q^{\mu} \rightarrow \Gamma_{\pi q}$. 

According to the discussions in sect. 4.2, in the static approximation 
the axial current
is given by the quark diagram, while for the pion absorption current we 
have
to consider also the exchange diagram (fig. 3). The expressions are as 
follows (see sect. 4.2):
\begin{eqnarray}
j_A^{\mu}(q)&=&\overline{\Gamma}_{N,P'} \int \frac{{\rm d}^4 k}{(2\pi)
^4} 
S_F(k+q) \Gamma_q^{\mu} S_F(k) \tau_s(p-k) \Gamma_{N,P} \label{qqs} \\  
j_{\pi}(q)&=&\overline{\Gamma}_{N,P'} \int \frac{{\rm d}^4 k}{(2\pi)^4} 
S_F(k+q) 
\Gamma_{\pi q} S_F(k) \tau_s(p-k) \Gamma_{N,P}\nonumber \\
&+& \frac{3}{M^2} \overline{\Gamma}_{N,P'} 2\,\Pi_N(P') \Gamma_{\pi q} 
\Pi_N(P) \Gamma_{N,P}, 
\label{qps} 
\end{eqnarray}
where the quark-diquark bubble graph is given by (see eq.(\ref{pin})
\begin{eqnarray}
\Pi_N(p)=- \int \frac{{\rm d}^4 k}{(2\pi)^4} S_F(p-k) \tau_s(k) \label
{bubble}
\end{eqnarray}
The Faddeev equation in the static approximation reduces to (see eq.
(\ref{Faddeev})
\begin{eqnarray}
\Pi_N(p) \Gamma_{N,P} = - \frac{M}{3} \Gamma_{N,P} \label{fes}
\end{eqnarray} 
and using (\ref{wtq}), (\ref{fes}) to calculate the divergence of the 
axial current (\ref{qqs})
we arrive at the PCAC relation (\ref{pcac}) in the static approximation.

\newpage

\begin{table}[hbt]
\begin{center}
\begin{tabular}{|c||c|c|c||c|}
\hline
case & I & II & III  & exp \\  \hline
$r_a$ & 0     & 0.25 & 0.66  & \\ \hline
$r_s$ & 0.73  & 0.63 & 0.50  & \\ \hline
$M_s [MeV]$ & 596   & 684 & 766 &   \\  \hline
$W_s$ [$\%$] & 100 & 93 & 61  & \\  \hline\hline
$\mu_p$ & 2.32 & 2.87 & 2.96 & 2.79 \\ \hline
$\mu_n$ & -1.39 & -2.08 & -2.62 & -1.91 \\  \hline
$g_A^{(3)}$  &  0.66 & 0.76 & 0.81 & 1.26 \\ \hline
$g_{\pi{\rm NN}}$ & 7.5 & 12.82 & 15.34 & 13.2 \\ \hline
$g_A^{(0)}$  & 0.60 &  0.41 & 0.30 & 0.2 $\sim$ 0.3 \\ \hline
\end{tabular}
\end{center}
\caption{The upper part of the table shows the diquark mass $M_s$ and 
the
contribution of the scalar diquark channel to the baryon number ($W_s$) 
for 
three different choices of the ratio $r_a$. For each case, the 
corresponding
value of $r_s$ is determined so as to reproduce the experimental 
nucleon mass.
The lower part of the table shows the static properties of the nucleon 
obtained
for these 3 cases in comparison to the experimental values.}  
\end{table}

\begin{table}[hbt]
\begin{tabular}{|c|c|c|c||c|c|c|}
\hline
case & I & II & III  & I & II & III \\  \hline
 & \multicolumn{3}{c||}{$\mu_p$ (exp.: 2.79)} & \multicolumn{3}{c|}
{$\mu_n$ (exp.: -1.91)} 
\\ \hline 
$Qs$         & 1.81 & 2.10 & 2.01 & -0.91 & -1.05 & -1.00  \\
$Ds$         & 0.04 & 0.04 & 0.01 & 0.04  & 0.04  &  0.01  \\
$Qa$         & 0    & 0    & 0    & 0     & -0.34 &  -0.83 \\
$Da$         & 0    & 0.05 & 0.18 & 0     & -0.02 &  -0.06 \\
$Dm$         & 0    & 0.12 & 0.12 & 0     & -0.12 &  -0.12 \\  \hline
sum          & 1.86 & 2.31 & 2.33 & -0.87 & -1.48 &  -2.00 \\
incl. $\pi$  & 2.32 & 2.87 & 2.92 & -1.39 & -2.08 &  -2.62 \\ \hline 
\hline 

case & I & II & III & I & II & III   \\  \hline

 & \multicolumn{3}{c||}{$g_A^{(3)}$ (exp.: 1.26)} & \multicolumn{3}{c|}
{$g_A^{(0)}$ 
(exp.: 0.2 $\sim$ 0.3)} 
\\ \hline 

$Qs$         & 0.77 & 0.55 & 0.24 & 0.77 & 0.55  & 0.24  \\
$Ds$         & 0    & 0    & 0    & 0    & 0     & 0  \\
$Qa$         & 0    & 0.03 & 0.03 & 0    & -0.09 & -0.09 \\
$Da$         & 0    & 0.03 & 0.17 & 0    & 0.05  & 0.26  \\
$Dm$         & 0    & 0.29 & 0.51 & 0    & 0     & 0 \\  \hline
sum          & 0.77 & 0.89 & 0.94 & 0.77 & 0.52  & 0.39 \\
incl. $\pi$  & 0.66 & 0.76 & 0.81 & 0.60 & 0.41  & 0.30 \\ \hline 
\end{tabular}
\begin{tabular}{|c|c|c|c|} 
\hline
case & I & II & III   \\  \hline

 & \multicolumn{3}{c|}{$g_{\pi NN}$ (exp.: 13.4)} 
\\ \hline 

$Qs$         & 12.00 & 14.14 & 13.61    \\
$Ds$         & 0     & 0     & 0        \\
$Qa$         & 0     & 0.28  & 0.34     \\
$Da$         & 0     & 0.58  & 2.56     \\
$Dm$         & 0     & -0.01 & -0.19    \\  
$Es$         & -2.89 & -2.77 & -1.74    \\
$Ea$         & 0     & -0.57 & -3.40    \\
$Em$         & 0     & 3.94  & 7.46     \\ \hline 
sum          & 9.12  & 15.58 & 18.64    \\
incl. $\pi$  & 7.50  & 12.82 & 15.35    \\ \hline
\end{tabular}

\caption{Static properties of the nucleon for the three cases I, II, 
III of
table 1. $Qs$ ($D_s$) denotes the contribution of the quark (diquark) 
diagram 
in the scalar channel, $Qa$ ($D_a$) refer to the quark (diquark) 
diagram in the a.v.
channel, and $Dm$ denotes the scalar-a.v. mixing in the diquark 
diagram. For
$g_{\pi NN}$ there are in addition the exchange diagram contributions
between the scalar channels ($Es$), between the a.v. channels ($Ea$), 
and the
mixing term ($Em$). The sum of these contributions, as well as the 
total sum 
including pion cloud effects, are shown.}  
\end{table}

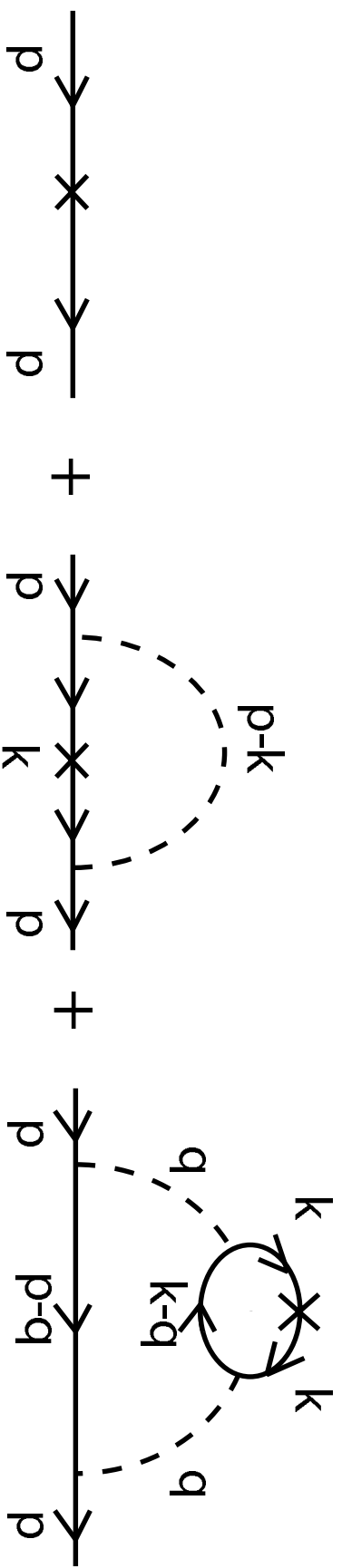
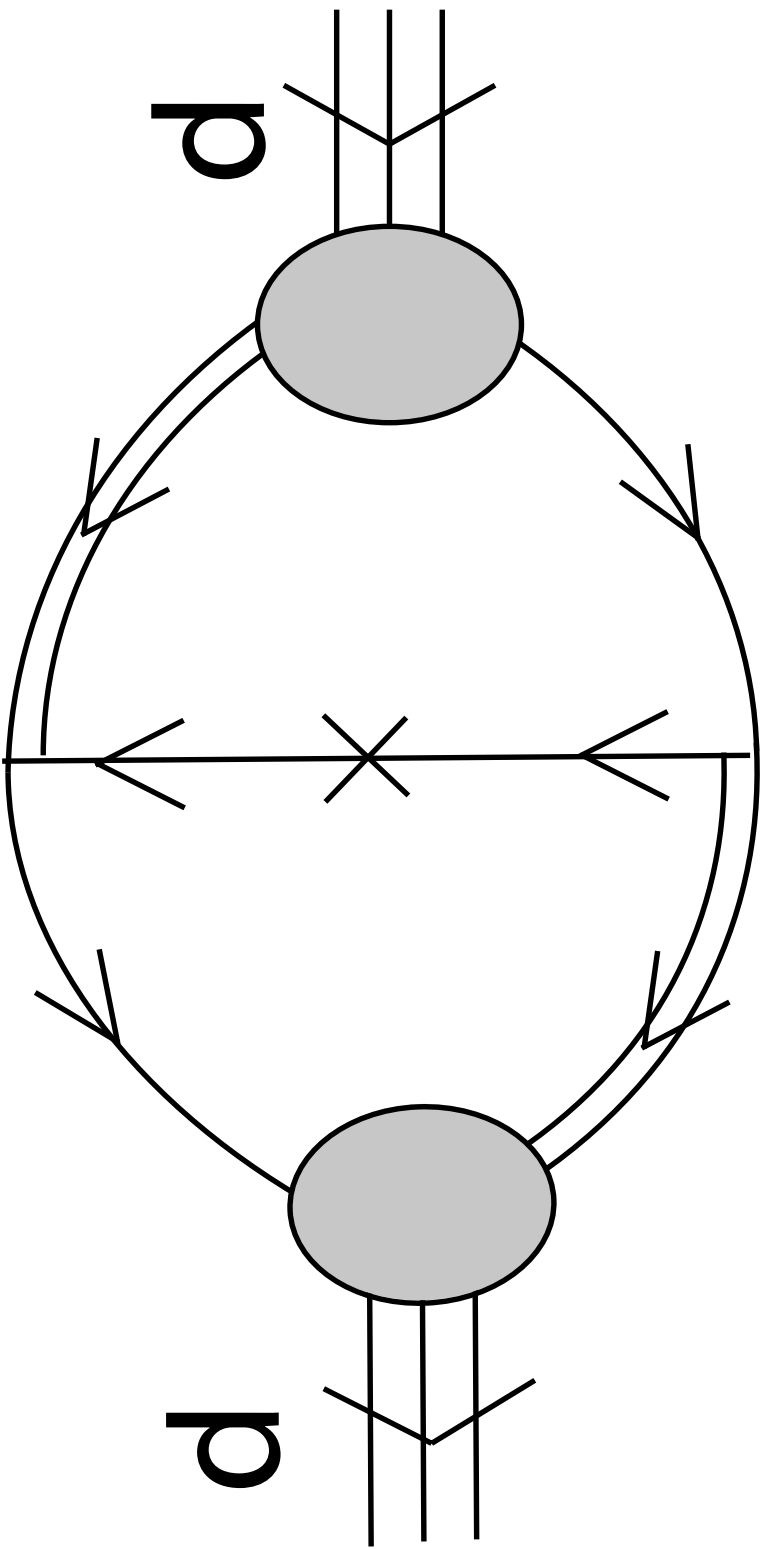
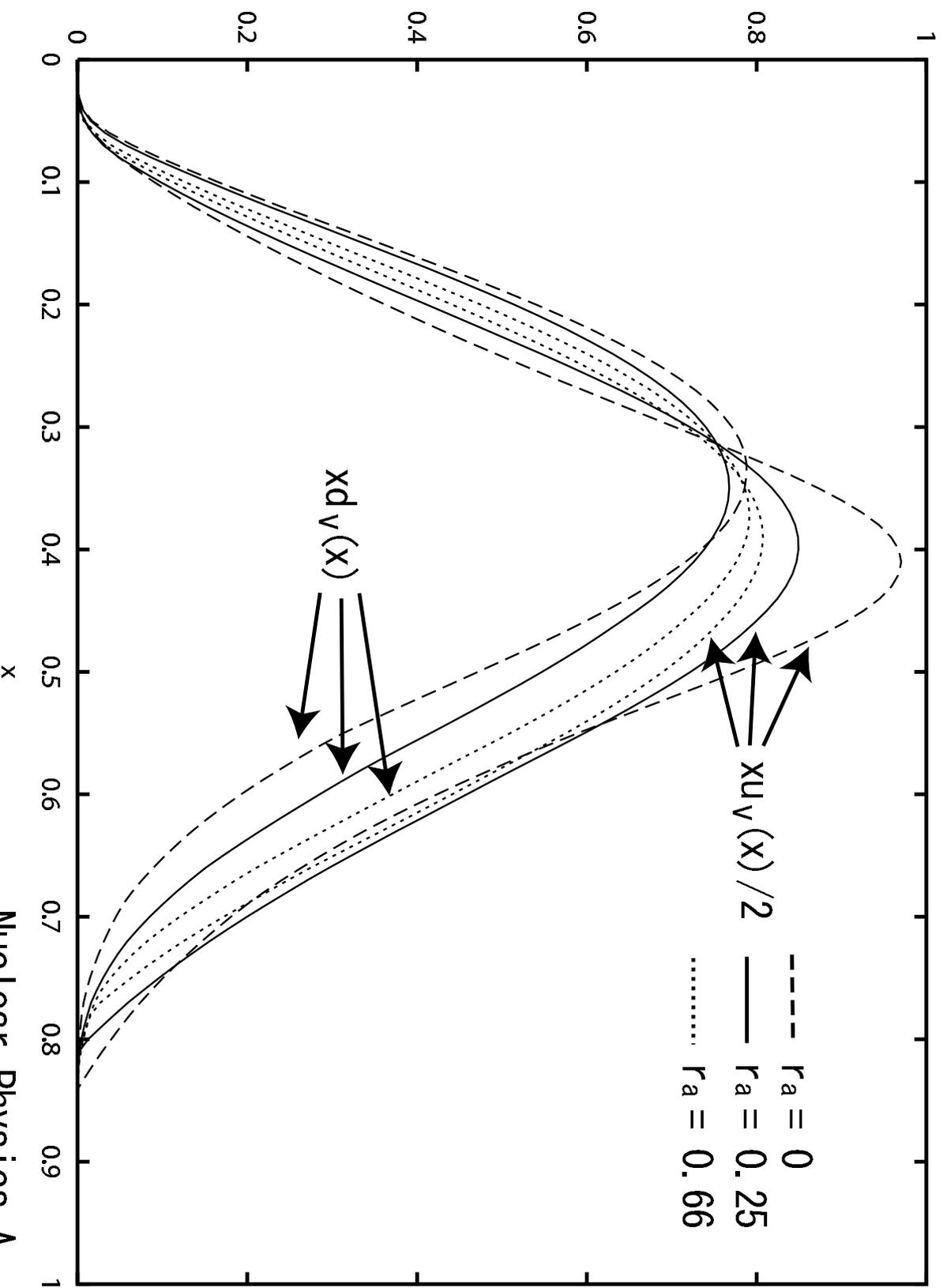
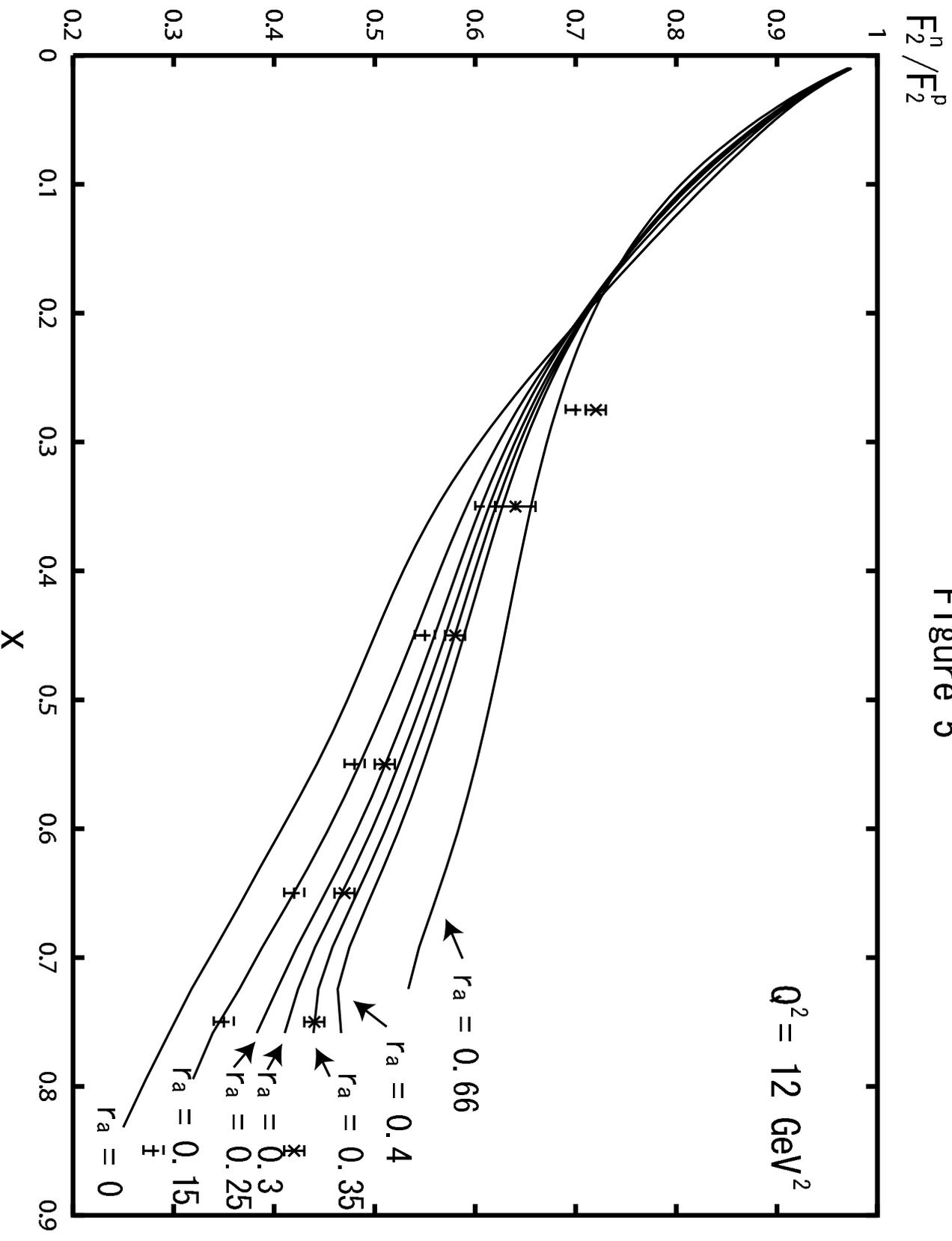
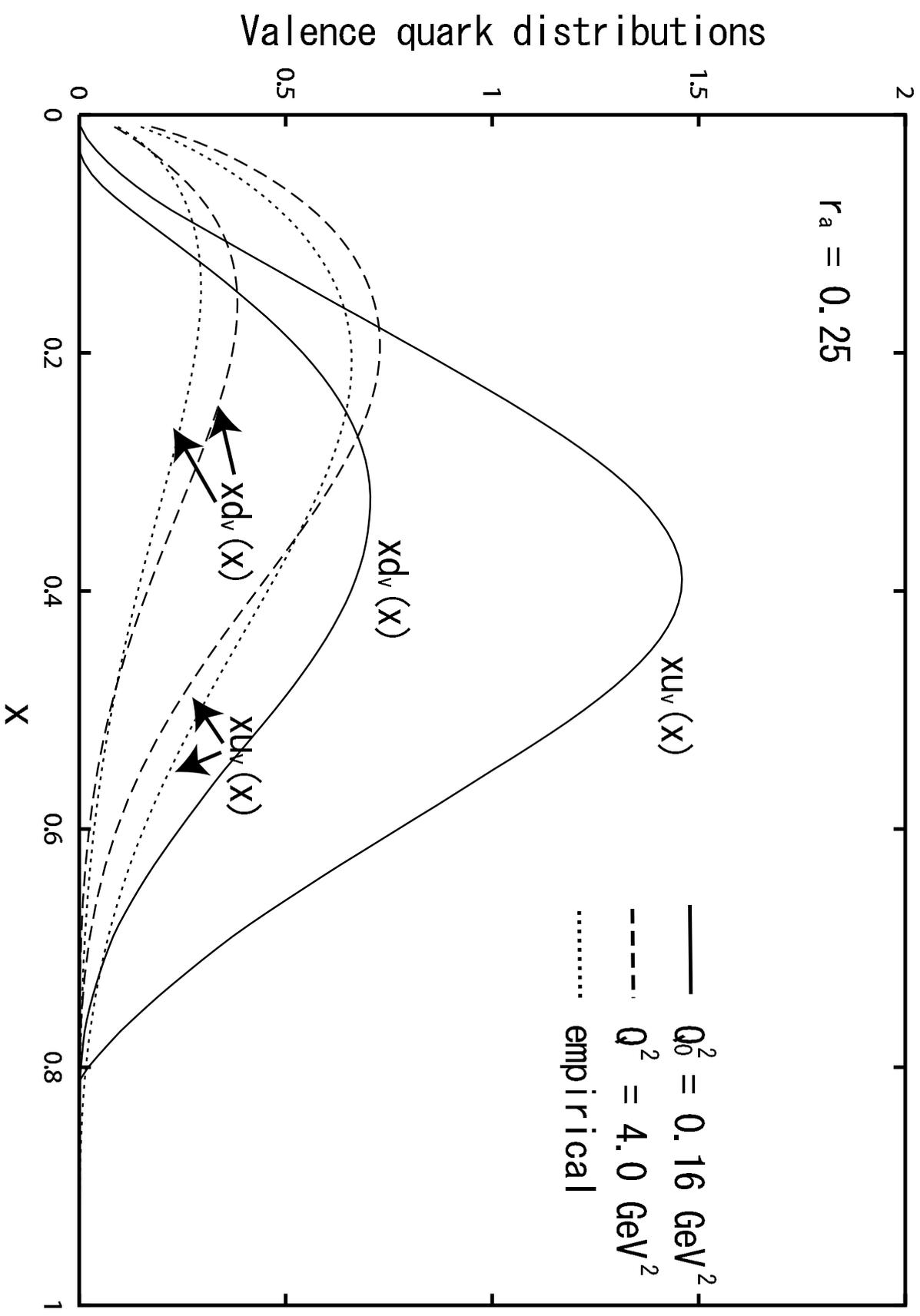
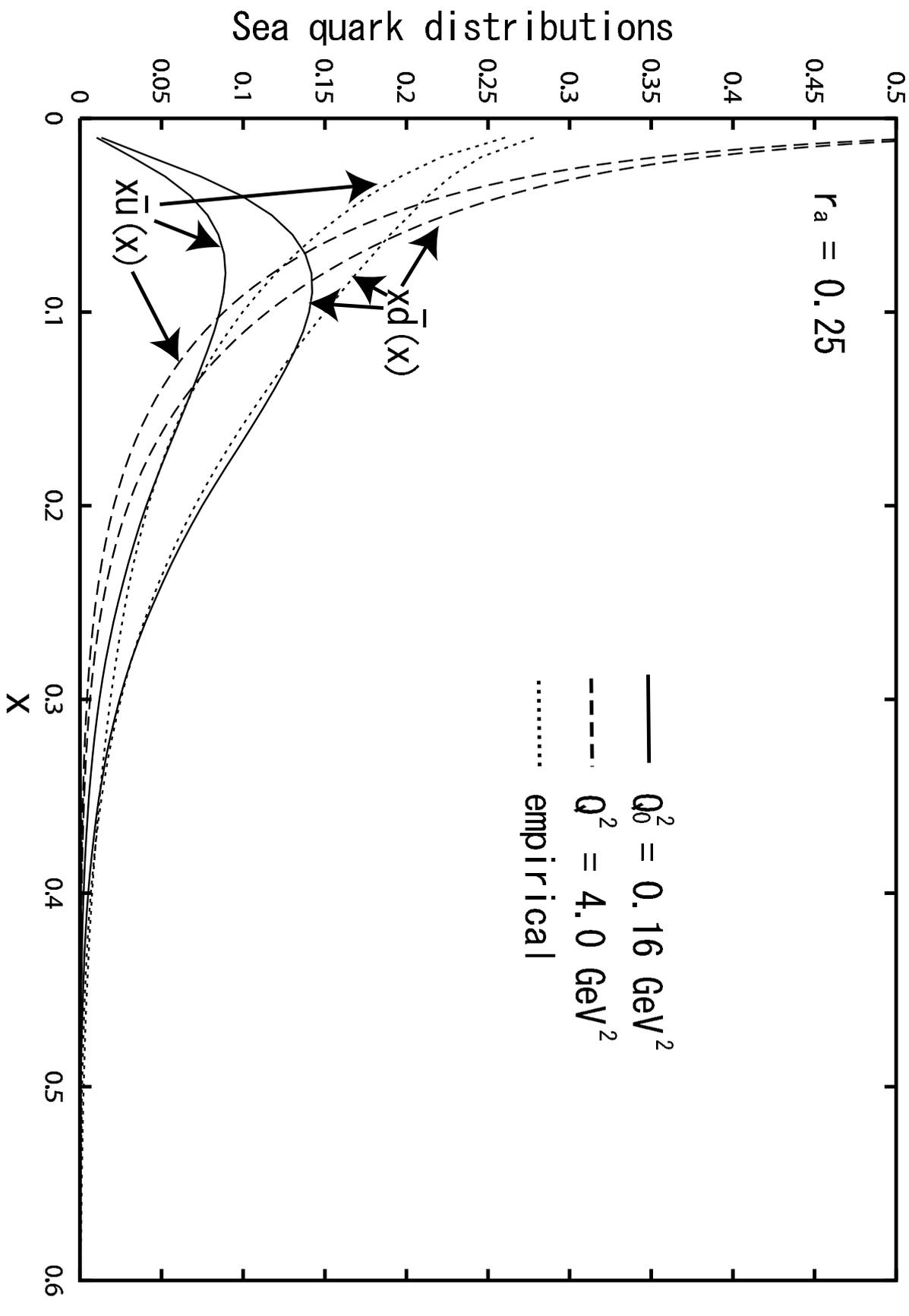
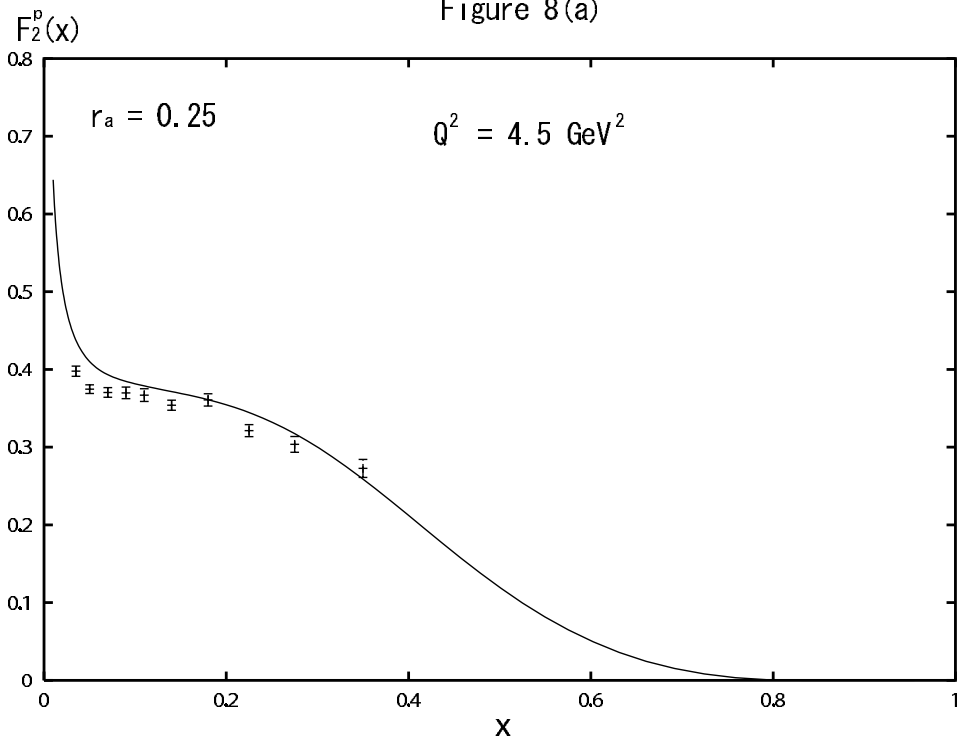
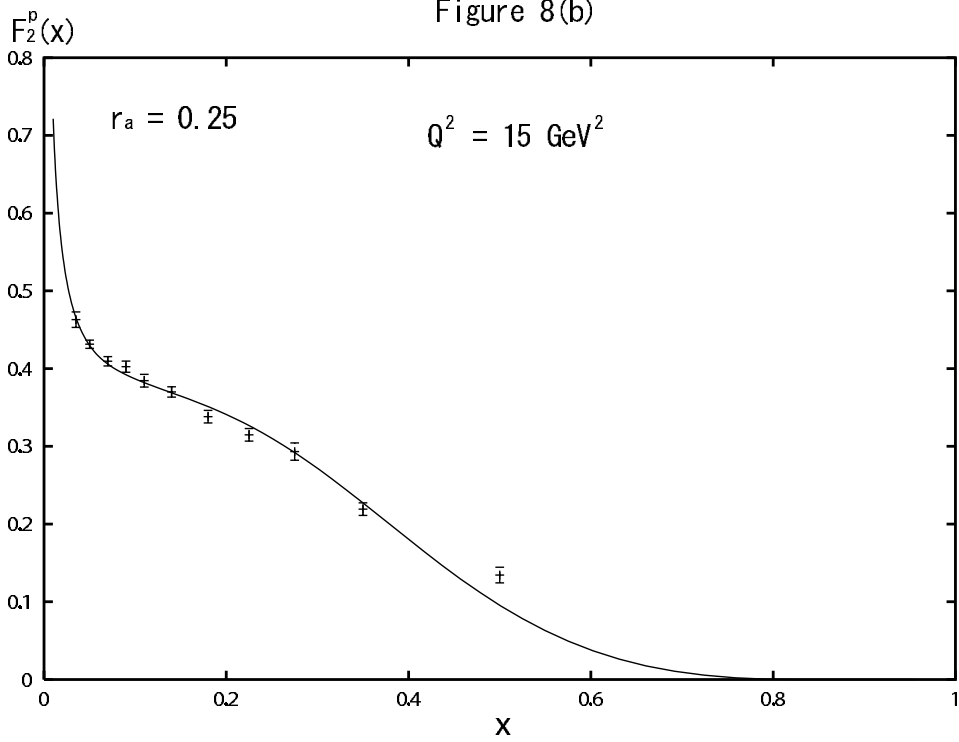
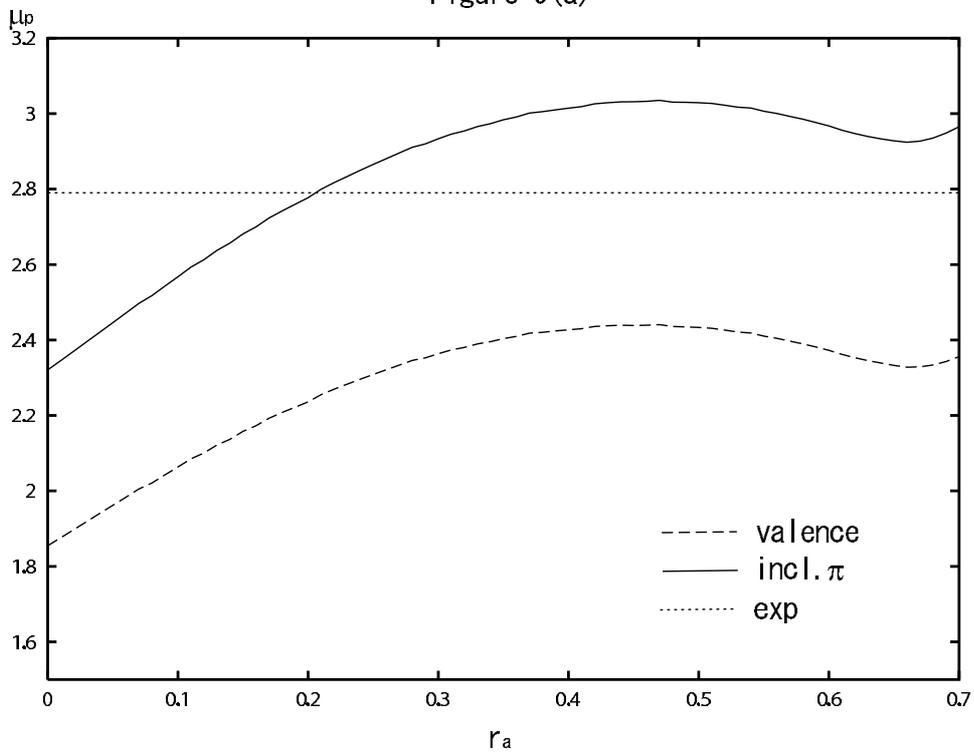
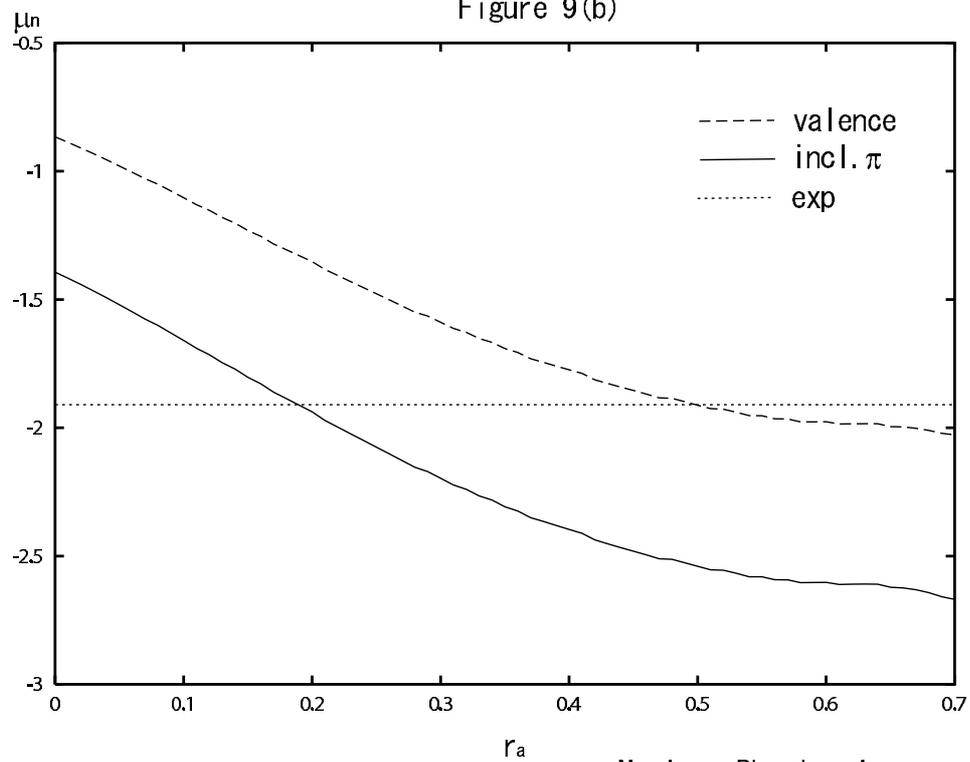
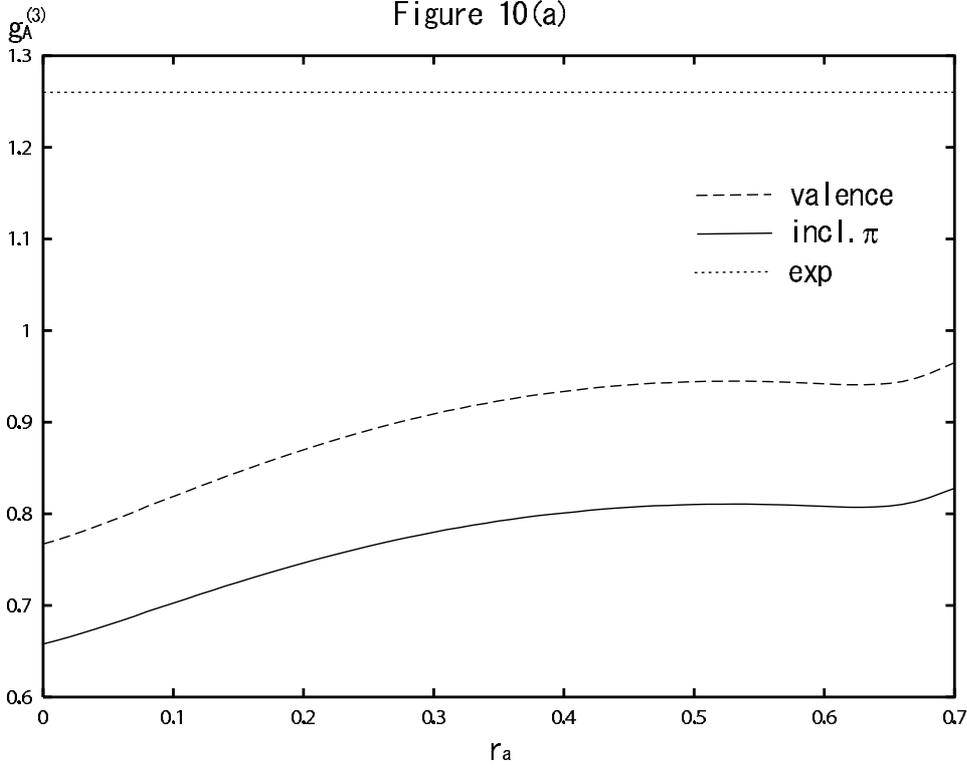
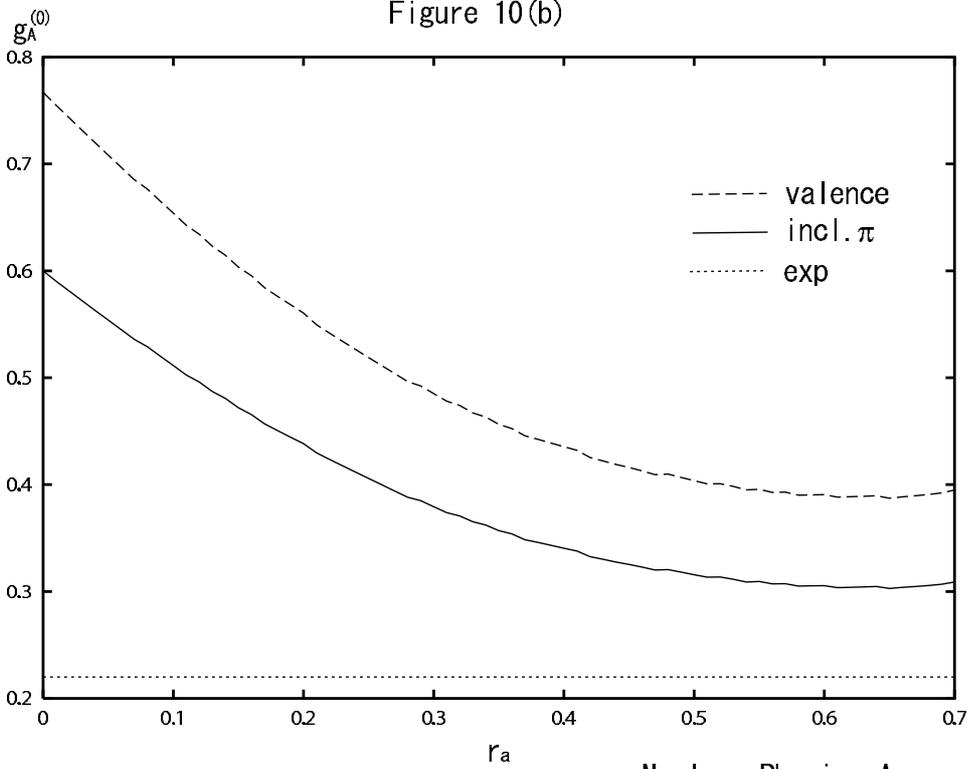
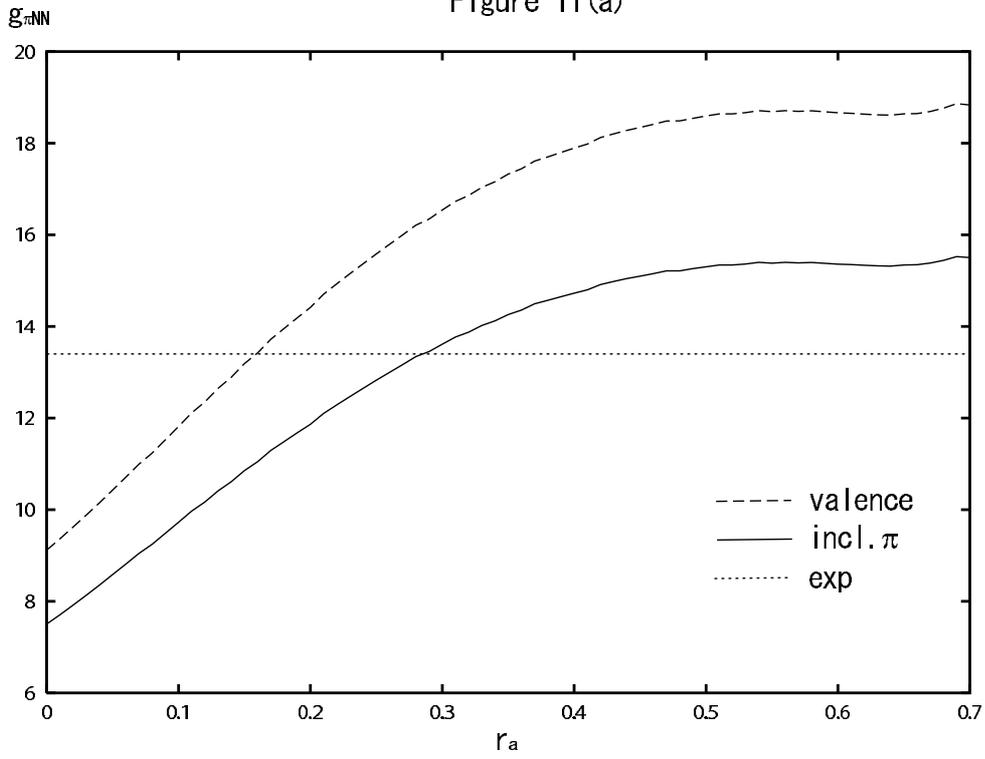
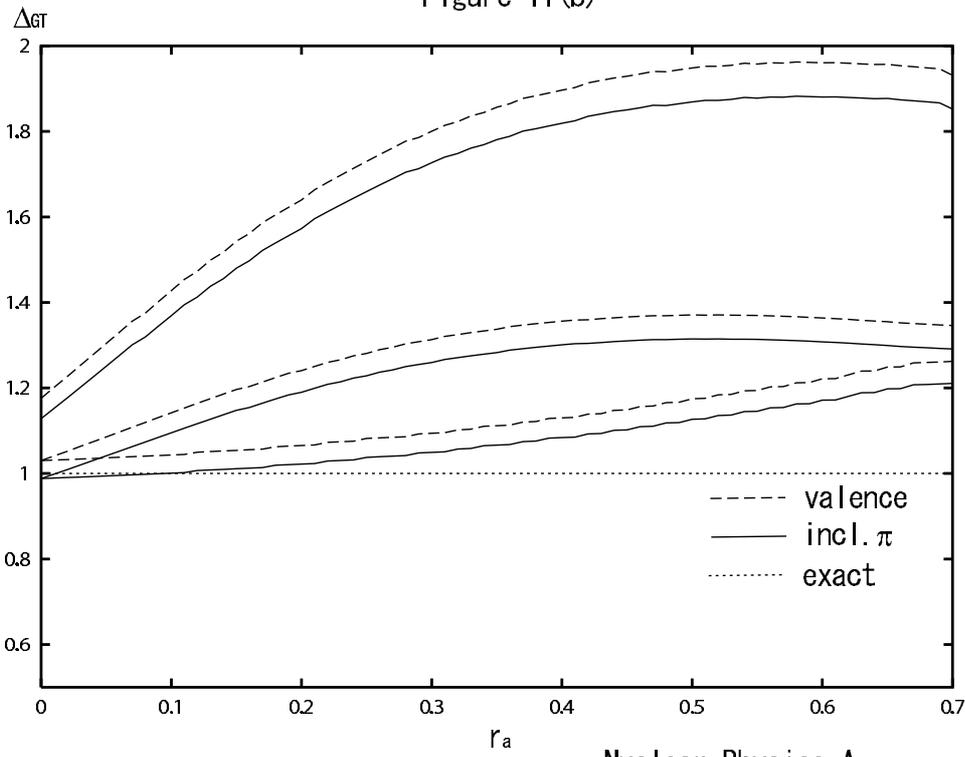
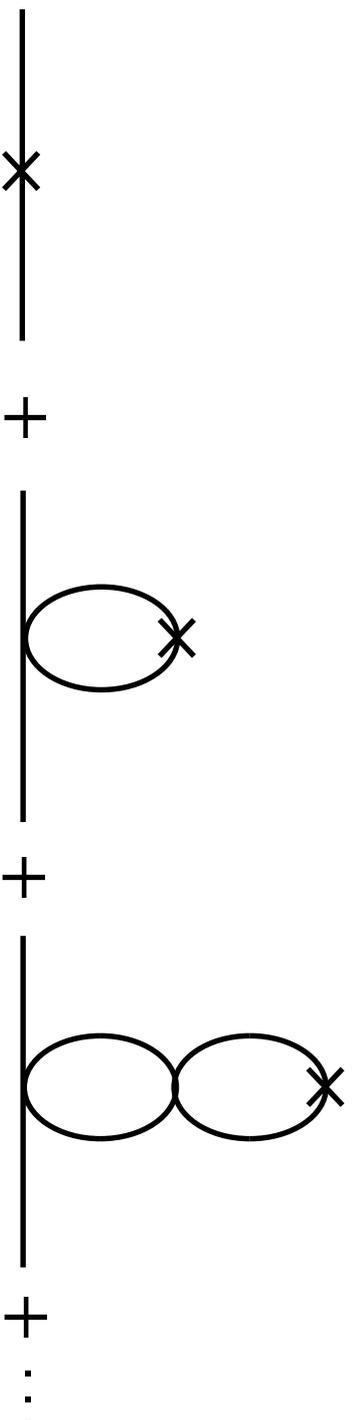
\end{document}